\PassOptionsToPackage{prologue,draft=false}{acmart} %
\documentclass[sigconf,anonymous=false]{acmart}
\newif\ifcamera
\camerafalse %
\newcommand{\appref}[1]{\ifcamera the full version\else Appendix~\ref{#1}\fi}
\def\noeditingmarks{} %

\AtBeginDocument{%
  }

\setcopyright{none}
\makeatletter
\renewcommand\footnotetextcopyrightpermission[1]{}
\makeatother
\acmConference[{}]{}{}{}
\acmBooktitle{}
\acmDOI{}
\acmISBN{}
\acmPrice{}

\usepackage{subcaption}
\usepackage{nicefrac}
\usepackage{booktabs}

\usepackage{balance}

\usepackage{pifont}

\usepackage{caption,subcaption}
\usepackage{graphicx}
\usepackage{booktabs}  %
\usepackage{dcolumn}   %
\usepackage{multirow}  %
\usepackage{makecell}  %
\usepackage{rotating}
\usepackage{xspace}
\usepackage{hyphenat}  %
\usepackage{color}
\usepackage{enumerate}
\usepackage{hyperref}
\usepackage{url}
\hypersetup{
    colorlinks=true,
    linkcolor=black,
    filecolor=magenta,      
    urlcolor=cyan,
    pdftitle={Overleaf Example},
    pdfpagemode=FullScreen,
    }

\usepackage[normalem]{ulem}      %

\usepackage{wrapfig}
\usepackage{textcomp}
\usepackage{tabularx}
\usepackage{pifont}
\usepackage{bm}
\usepackage{adjustbox}
\usepackage{blindtext}
\usepackage{ifthen}

\makeatletter
\def\NAT@def@citea{\def\@citea{\NAT@separator}}
\makeatother

\usepackage{float}
\usepackage{newfloat}
\DeclareFloatingEnvironment[
    fileext=los,
    listname=List of Schemes,
    name=Scheme,
    placement=H,
    within=section,
]{scheme}

\usepackage{wrapfig}

\usepackage{balance}

\usepackage{colortbl} %
\usepackage{array}    %

\usepackage{dblfloatfix}

\usepackage{listings}
\definecolor{maroon}{cmyk}{0, 0.87, 0.68, 0.32}
\definecolor{halfgray}{gray}{0.55}
\definecolor{ipython_frame}{RGB}{207, 207, 207}
\definecolor{ipython_bg}{RGB}{247, 247, 247}
\definecolor{ipython_red}{RGB}{186, 33, 33}
\definecolor{ipython_green}{RGB}{0, 128, 0}
\definecolor{ipython_cyan}{RGB}{64, 128, 128}
\definecolor{ipython_purple}{RGB}{170, 34, 255}
\definecolor{syellow}{HTML}{B58900}
\definecolor{sorange}{HTML}{CB4B16}
\definecolor{smagenta}{HTML}{D33682}
\definecolor{sviolet}{HTML}{6C71C4}
\definecolor{sblue}{HTML}{268BD2}
\definecolor{scyan}{HTML}{2AA198}
\definecolor{sgreen}{HTML}{859900}

\usepackage{algpseudocode}
\algrenewcomment[1]{\hfill// #1}%
\algnotext{EndFunction}
\algnotext{EndFor}
\algnotext{EndWhile}
\algnotext{EndIf}

\usepackage{amsmath,amscd}
\usepackage{amsfonts}
\usepackage{amsthm}
\usepackage[most]{tcolorbox}
\tcbuselibrary{skins,breakable}
\usepackage{mathbbol}
\usepackage{lipsum}
\algrenewcommand\algorithmicindent{1em}
\algrenewcommand{\algorithmiccomment}[1]{{\color{teal}\hfill// #1}}
\usepackage{enumitem}

\newif\ifextended
\newif\iflongbatching
\newif\ifsubmission
\newif\ifelementary

\ifx\buildextended\undefined
\else
    \extendedtrue
\fi

\ifx\noeditingmarks\undefined
\newcommand{\pgwrapper}[3]{\begingroup \color{#1} #2: #3 \endgroup}
\newcommand{\pgwrapperb}[1]{\textbf{#1}}
\else
   \newcommand{\pgwrapperb}[1]{}
   \newcommand{\pgwrapper}[3]{}
\fi

\newcommand{\todo}[1]{\pgwrapper{red}{TODO}{#1}}

\def\hn{\usefont{OT1}{phv}{mc}{n}\selectfont}

\newcommand{\mpfont}{\hn\scriptsize}

\ifx\noeditingmarks\undefined
\newcommand{\MPworker}[2]{{\color{#1}\vrule\vrule}{\marginpar{\color{#1}\mpfont #2}}}
\else
    \newcommand{\MPworker}[2]{}
\fi

\ifx\noeditingmarks\undefined
    
\else
    
\fi

\newcommand{\sys}{Armadillo\xspace}

\theoremstyle{definition}
\newtheorem{theorem}{Theorem}
\newtheorem{definition}{Definition}

\newtheorem{remark}{Remark}
\newtheorem{lemma}{Lemma}

\makeatletter
\renewcommand*{\@fnsymbol}[1]{\ensuremath{\ifcase#1\or \star\or \dagger\or \ddagger\or
   \mathsection\or \mathparagraph\or \|\or **\or \dagger\dagger
   \or \ddagger\ddagger \else\@ctrerr\fi}}
\makeatother

\newcommand{\Z}{\mathbb{Z}}

\newcommand{\floor}[1]{\lfloor#1\rfloor}

\newcommand{\etal}{et al.\xspace}

\makeatletter
\def\imod#1{\allowbreak\mkern10mu({\operator@font mod}\,\,#1)}
\makeatother


\def\compactify{\itemsep=0in \topsep=2pt \parsep=0.00in \partopsep=0pt
\leftmargin=2em}
\let\latexusecounter=\usecounter

  {\begin{list}{\labelitemi}{\itemsep3pt \topsep3pt \parsep0.00in
  \partopsep=3pt \leftmargin1.2em}}%
  {\end{list}}
\newenvironment{myitemize2}%
  {\begin{list}{\labelitemi}{\itemsep1pt \topsep2pt \parsep0.00in
  \partopsep=1pt \leftmargin1.2em}}%
  {\end{list}}
  {\begin{list}{\labelitemi}{\itemsep2pt \topsep2pt \parsep0.00in
  \partopsep=0pt \leftmargin1.2em}}%
  {\end{list}}
  {\begin{list}{\threequartdash}{\itemsep3pt \topsep3pt \parsep0.00in
  \partopsep=3pt \leftmargin1.5em}}%
  {\end{list}}

\def\compactsortof{\itemsep=0in \topsep=2pt \parsep=0.00in \partopsep=0pt
\leftmargin=1.7em}
\newenvironment{myenumerate2}
  {\def\usecounter{\compactsortof\latexusecounter}
   \begin{enumerate}}
  {\end{enumerate}\let\usecounter=\latexusecounter}

\def\compactsqueeze{\itemsep=0pt \topsep0pt \parsep=0ex \partopsep=0pt
\leftmargin=1.63em}

\ifx\normalpar\undefined
  
\else
  
\fi

\def\discretionaryslash{\discretionary{/}{}{/}}
{\catcode`\/\active
\gdef\URLprepare{\catcode`\/\active\let/\discretionaryslash
        \def~{\char`\~}}}%
\def\URL{\bgroup\URLprepare\realURL}%
\def\realURL#1{\tt #1\egroup}%

\begin{document}

\title{\sys{}: Robust Single-Server Secure Aggregation for Federated Learning with Input Validation}

\author{Yiping Ma}
\authornote{Work done at University of Pennsylvania and during an internship at JPMorgan AI Research.}
\email{yipingma@berkeley.edu}
\orcid{0000-0002-3183-9402}
\affiliation{%
  \institution{University of Pennsylvania and UC Berkeley}
  \city{Philadelphia}
  \country{USA}
}
\author{Yue Guo}
\email{yue.guo@jpmchase.com}
\orcid{0000-0002-0712-6656}
\affiliation{%
  \institution{J.P. Morgan AI Research and AlgoCRYPT CoE}
  \city{New York}
  \country{USA}}
\author{Harish Karthikeyan}
\email{harish.karthikeyan@jpmchase.com}
\orcid{0000-0002-1787-4906}
\affiliation{%
  \institution{J.P. Morgan AI Research and AlgoCRYPT CoE}
  \city{New York}
  \country{USA}}
\author{Antigoni Polychroniadou}
\email{antigoni.polychroniadou@jpmorgan.com}
\orcid{0009-0003-0125-2971}
\affiliation{%
  \institution{J.P. Morgan AI Research and AlgoCRYPT CoE}
  \city{New York}
  \country{USA}}
\email{}

\renewcommand{\shortauthors}{Ma et al.}

\newcommand{\G}{\mathbb{G}}
\newcommand{\bH}{\mathbb{H}}

\renewcommand{\vec}[1]{\mathbf{#1}}

\newcommand{\antig}[1]{
  \todo[inline,color=green!20]{\textbf{Antigoni:} #1}
}

\newcommand{\Poseidon}{\textsc{Poseidon}}
\newcommand{\Commit}{\mathrm{Commit}}
\newcommand{\MAC}{\mathrm{MAC}}

\newcommand{\client}{\ensuremath{\testsf{C}}}
\newcommand{\server}{\ensuremath{\textsf{S}}}

\renewcommand{\bold}[1]{{\mbox{$\mathbf{#1}$}}}
\newcommand{\bdx}{\bold{x}}
\newcommand{\bdy}{\bold{y}}
\newcommand{\bdz}{\bold{z}}
\newcommand{\bdw}{\bold{w}}
\newcommand{\bdv}{\bold{v}}
\newcommand{\bdd}{\bold{d}}
\newcommand{\bda}{\bold{a}}
\newcommand{\bdA}{\bold{A}}
\newcommand{\bdb}{\bold{b}}
\newcommand{\bdB}{\bold{B}}
\newcommand{\bdc}{\bold{c}}
\newcommand{\bdC}{\bold{C}}
\newcommand{\bde}{\bold{e}}
\newcommand{\bdE}{\bold{E}}
\newcommand{\bds}{\bold{s}}
\newcommand{\bdk}{\bold{k}}
\newcommand{\bdS}{\bold{S}}
\newcommand{\bdr}{\bold{r}}
\newcommand{\bdR}{\bold{R}}
\newcommand{\bdu}{\bold{u}}
\newcommand{\bdG}{\bold{G}}
\newcommand{\bdM}{\bold{M}}
\newcommand{\bfs}{\bold{s}}
\newcommand{\bfw}{\bold{w}}
\newcommand{\ecq}{Q}

\newcommand{\calD}{\mathcal{D}}
\newcommand{\calS}{\mathcal{S}}
\newcommand{\calX}{\mathcal{X}}
\newcommand{\calB}{\mathcal{B}}

\newcommand{\poly}{\ensuremath{\mathsf{poly}}\xspace}
\newcommand{\polylog}{\ensuremath{\mathsf{polylog}}\xspace}

\newcommand{\FF}{\mathbb{F}}
\newcommand{\PRG}{\texttt{PRG}}

\newcommand{\lwesecret}{\bold{s}}

\newcommand{\nlookup}{\mathsf{nlookup}\xspace}

\newcommand{\Enc}{\mathsf{Enc}}
\newcommand{\Dec}{\mathsf{Dec}}

\newcommand{\prover}{\mathcal{P}}
\newcommand{\verifier}{\mathcal{V}}
\newcommand{\ZKP}{\mathsf{ZKP}}
\newcommand{\zkio}{\mathsf{io}}
\newcommand{\pp}{\mathsf{pp}}
\newcommand{\witness}{\mathsf{wt}}
\newcommand{\zkstate}{\mathsf{st}}

\newcommand{\ChooseSet}{\textsc{ChooseSet}}

\newcommand{\Encode}{\mathsf{Encode}}
\newcommand{\Decode}{\mathsf{Decode}}

\newcommand{\AsymEnc}{\mathsf{AsymEnc}}
\newcommand{\AsymDec}{\mathsf{AsymDec}}
\newcommand{\AsymGen}{\mathsf{AsymGen}}

\newcommand{\SymGen}{\mathsf{SymAuthGen}}
\newcommand{\SymEnc}{\mathsf{SymAuthEnc}}
\newcommand{\SymDec}{\mathsf{SymAuthDec}}

\newcommand{\PackedSS}{\mathsf{PackedSS}}

\newcommand{\Share}{\emph{\mathsf{Share}}\xspace}
\newcommand{\Recon}{\emph{\mathsf{Recon}}\xspace}
\newcommand{\secret}{\rho}
\newcommand{\share}{\secret}
\newcommand{\polynomial}{p}

\newcommand{\VSS}{\textrm{VSS}}
\newcommand{\Shamir}{\textrm{Shamir}}
\newcommand{\CommitPoly}{\textrm{CommitPoly}}
\newcommand{\ShareAndCommit}{\textrm{Share-Commit}}
\newcommand{\Verify}{\textrm{Verify}}
\newcommand{\QUAL}{\textrm{QUAL}}

\newcommand{\commit}{\mathsf{com}}

\newcommand{\Fcomp}{\mathcal{F}}
\newcommand{\Fsemi}{\mathcal{F}_{\text{semi}}}
\newcommand{\Fmal}{\mathcal{F}_{\text{mal}}}
\newcommand{\Fmalrb}{\mathcal{F}_{\text{mal-robust}}}
\newcommand{\Fsetup}{\mathcal{F}_{\text{setup}}}
\newcommand{\Frand}{\mathcal{F}_{\text{rand}}}
\newcommand{\Fsumt}{\mathcal{F}_{\text{sum}}}
\newcommand{\REAL}{\textrm{REAL}}
\newcommand{\IDEAL}{\textrm{IDEAL}}
\newcommand{\Hybrid}{\textrm{Hybrid}}
\newcommand{\Sim}{\mathcal{S}}
\newcommand{\View}{\textrm{View}}
\newcommand{\Adv}{\mathsf{Adv}}

\newcommand{\ndec}{D}
\newcommand{\Coms}{Decryptors\xspace}
\newcommand{\coms}{decryptors\xspace}
\newcommand{\Com}{Decryptor\xspace}
\newcommand{\com}{decryptor\xspace} 
\newcommand{\comset}{\mathcal{C}}

\newcommand{\CSEnc}{\mathbb{CS}_{\text{enc}}}
\newcommand{\CSShare}{\mathbb{CS}_{\text{shares}}}
\newcommand{\CSInput}{\mathbb{CS}_{\text{input}}}
\newcommand{\Circuit}{\mathsf{C}}

\newcommand{\Hash}{\ensuremath{\mathsf{Hash}}\xspace}
\newcommand{\Test}{\ensuremath{\mathsf{Test}}\xspace}

\newcommand{\online}{\ensuremath{\mathcal{O}}}

\newcommand{\cmark}{\ding{51}} %
\newcommand{\xmark}{\ding{55}} %
\begin{abstract}
This paper presents a secure aggregation system \sys{} that has disruptive
resistance against adversarial clients, such that any coalition of malicious
clients can affect the aggregation result only by misreporting their private
inputs in a pre-defined legitimate range. \sys{} is designed for federated
learning setting, where a single powerful server interacts with many weak
clients iteratively to train models on client's private data. While a few prior
works consider disruption resistance under such setting, for an aggregation on
$n$ clients they either require high cost per client (Chowdhury et al. CCS '22)
or concretely many rounds that is logarithmic in $n$ (Bell et al. USENIX
Security '23). Although disruption resistance can be achieved generically with
zero-knowledge proof techniques (which we also use in this paper), we realize an
efficient system with two new designs: 1) a simple two-layer secure aggregation
protocol that requires only simple arithmetic computation; 2) an agreement
protocol that removes the effect of malicious clients from the aggregation with
low round complexity. With these techniques, \sys{} runs in 3 rounds per
aggregation (our round complexity is independent of $n$) with computationally
lightweight server and clients.  

\end{abstract}

\settopmatter{printfolios=true, printccs=false, printacmref=false}
\maketitle
\pagestyle{plain}
\thispagestyle{plain}

\section{Introduction}\label{s:intro}

Can a server aggregate private data from many clients without learning any
  individual's data? And can this be done when clients {\em arbitrarily disrupt} the
  aggregation? In this work, we provide an affirmative answer to these questions
  with an aggregation system with strong privacy and robustness guarantees, 
  even when we only have a single untrusted server coordinating the aggregation.

While the secure aggregation problem is decades old, there is a renewed interest
  motivated by the emergence of federated learning~\cite{mcmahan17communication}: such
  a setting has a particular communication model and system constraints---{\em a central
  powerful server interacts with many weak clients}.
Here, ``powerful'' vs. ``weak'' is with respect to computation power, communication
  bandwidth, and availability. Each client has a locally trained model (which can be
  viewed as a high-dimensional vector), and the server should learn the sum of
  these models (the sum of the vectors component-wise) but nothing else. 

This setting makes secure aggregation challenging due to the high likelihood of 
  adversarial clients in a large participant pool. 
Many existing aggregation systems~\cite{AddankiGJOP22, corrigan-gibbs17prio} 
  that tolerates disruption assumes a strictly weaker trust model than the
  model we consider in this work. 
They require two or more non-colluding servers with at least one being trusted to achieve their goals, yet in real-world federated learning deployment, the single-server architecture seems to be the de facto choice: an organization that runs the training either internally 
  operate their own servers while ensuring isolation among the servers, which is rarely realistic; or 
  they set up external servers elsewhere, which introduces significant engineering overhead and operational risks. 
  And indeed, industry precedent has consistently relied on the single-server
  model~\cite{mcmahan17communication, bonawitz17practical, hartmann24distributed}. 
A few prior single-server aggregation solutions that tolerate disruption, unfortunately have high costs: 
  Chowdhury et al.~\cite{CCS:CGJM22} has a per-client computational workload
  quadratic in the number of clients, 
  and Bell et al.~\cite{bell22acorn} despite its modest client cost has too many rounds 
  to be efficient---a single aggregation takes logarithmic rounds in the number
  of clients and concretely 9--13 rounds for 1K clients.
Some other solutions~\cite{bell22acorn, lycklama23rofl} settle for a 
  relatively weak guarantee where the aggregation has to 
  abort once disruption is detected. 
That is, there was essentially no
  affordable disruptive-resistant secure aggregation schemes under practical adversarial models.

This gap drives our work: our view is that ensuring privacy for clients while
  resisting disruption can be achieved in only 3 rounds,
  even in the presence
  of an honest-but-curious server colluding with a subset of malicious clients
  below a certain threshold. Our system \sys{} guarantees the following properties: 1)
  privacy, i.e., the server at most {\em learns the sum of inputs} from clients
  but nothing else, 2) robustness, i.e., the server, if following our protocol,
  is {\em ensured to get the sum} regardless how clients passively drop out or
  actively disrupt the aggregation. So far, a malicious client could still
  disrupt the result with an invalid input, but we show that \sys{} can seamlessly integrate with
  existing efficient input validation techniques~\cite{GHL22}, resulting in a complete
  disruption-resistant system.
Our reduction on round complexity does not come with a price on computational time: even when integrated with 
  input validation techniques, the concrete computation stays close to that of
  the best prior work ACORN-robust, a protocol in Bell et al.~\cite{bell22acorn}:
  about $1.6\times$ its client computation, and less server computation.
Varying the number of clients and the fraction of adversarial clients,
\sys{} outperforms ACORN-robust by 3--4.3$\times$ in rounds.

To achieve disruption resistance, \sys{} uses a generic paradigm: take a secure
  aggregation protocol, every client proves that every step of its execution has
  been done correctly and the server verifies the proof. The core challenge is
  to make this efficient because cryptographic proofs are expensive; in fact,
  this is more challenging than it may seem---most of the prior
  works~\cite{lycklama23rofl, bell22acorn, CCS:CGJM22} making strides toward
  robustness do not follow this paradigm. Two key design ideas help us achieve
  lightweight clients and server. First, we design a secure aggregation protocol
  in which the bulk of computation is a simple linear computation, and
  importantly, it is sufficient to get robustness as long as the clients prove
  the correctness of the linear part (which is computationally efficient). Then,
  we structure all these proof statements (together with the input validation)
  as a single inner-product relation, so that with existing proof
  systems~\cite{bunz18bulletproof}, the server can batch verify the $n$ clients'
  proofs with shared multi-exponentiations, so that each additional client
  costs a few milliseconds.

\subsection{Technical contributions}

\paragraph{Efficiency improvement over prior works.}
In Figure~\ref{fig:comparison}, we compare the asymptotic cost of \sys{} 
  with prior works with similar properties. 
  The only prior work that achieves the same property as \sys{} and with
  reasonable per client cost is
  ACORN-robust~\cite{bell22acorn}, in which an aggregation on $n$ clients takes
  $6 + O(\log n)$ communication rounds between server and all the clients.
\sys{} has 3 rounds regardless of the number of clients and the
  asymptotic computation and communication costs on par with the best prior result.
As evidenced in prior works~\cite{ma23flamingo}, round complexity is critical for
  end-to-end run time in the federated learning setting. In each
  communication round, the server has to wait until
  it receives a desired proportion of the clients' responses. If a protocol has
  many rounds, the time spent on waiting may dominate over the actual
  computation cost.

In various circumstances, we have 3--4.3$\times$ 
  improvement on the number of communication rounds, and up to $8.9\times$
  improvement on the run time for computing a sum. Our competitive advantage over
  ACORN-robust becomes more significant with a higher corruption rate among
  clients. These concrete improvements are detailed in
  Section~\ref{s:eval}.

\paragraph{Technical novelty.}
We utilize a key-and-message homomorphic encryption scheme to build a simple
two-layer secure aggregation protocol: the clients first send encryption of
their inputs, and the server sums up these ciphertexts (the {\em outer} layer).
Then, it runs another aggregation on the keys to decrypt the sum of the
ciphertexts (the {\em inner} layer). Note that the keys are much smaller than
the inputs in federated learning; this two-layer paradigm reduces our original
problem to a secure aggregation with smaller inputs. While the tool of
key-and-message homomorphism has appeared in the secure aggregation
context~\cite{liu2022sash,li23lerna, bell22acorn, stevens22efficient},
our design is concretely different from prior works, and we address the
robustness challenge (that almost none of these works have) by adding only
lightweight components to the aggregation protocol.

We instantiate both the outer and inner layers with primarily linear
computation; this allows the clients to generate low-cost proof of computation.
We use commitment to bind the two layers, ensuring that the key underlying the
ciphertext in the outer layer is indeed the input to the inner layer. Crucially,
all these must be done in a concretely feasible way for computationally weak
clients. Our technical overview is provided in Section~\ref{s:overview}.

\paragraph{Formal security guarantees.}
We provide a simulation-based proof of privacy against an honest-but-curious
server that colludes with a subset of clients, which may deviate arbitrarily
from the protocol (\S\ref{s:pb}); ACORN-robust~\cite{bell22acorn} also assumes a
semi-honest server for privacy. We also show that \sys{} has
robustness (Def.\ref{def:robustness}) against any coalition of up to an
$\eta$ fraction of adversarial clients,  
and combined with existing input validation techniques, we get disruption
resistance (Def.\ref{def:disruption-resistance}).
While some existing protocols~\cite{bell22acorn, lycklama23rofl} claimed
  privacy against a malicious server, these protocols have not been proven secure
  when adding in their input validation techniques; some of them indeed do not
  include a security proof.
We argue that any secure aggregation protocol with a security proof
  relying on the programmability of the random oracle model cannot incorporate input validation techniques
  that require proving the correct computation of a later programmed value.
We discuss this in detail in Section~\ref{s:security}.

\subsection{Problem statement}\label{s:pb}

\begin{figure*}[!tb]

\centering
\resizebox{0.95\textwidth}{!}{\begin{tabular}{@{}llllllll@{}}
\toprule
& Client comm.     & Client comp.      & Server comm.     & Server comp.  & Rounds & Robustness & Input Val.   \\                %
\midrule
Effiel~\cite{CCS:CGJM22} & $\ell n^2$ & $\ell n^2$ & $\ell n^3$ & $\ell n^3$ & 4 & $\checkmark$ & Generic  \\       %
\midrule
RoFL~\cite{lycklama23rofl} & $\ell + \log n$ & $\ell \log n $ & $ \ell n + n \log n $  & $\ell n$ & 6 & $\times$ & $L_2, L_\infty$ \\ %
\midrule
ACORN-detect~\cite{bell22acorn} & $\ell+\log n$   & $\ell \log n $  & $\ell n + n\log n$    & $ \ell n $  & 7  & $\times$ & $L_2, L_\infty$   \\ %
\midrule
ACORN-robust~\cite{bell22acorn} & $\ell +\log^2 n$ & $\ell \log n + \log^2 n$  & $\ell n  + n \log^2 n$      & $\ell n + n \log^2 n  $  & $6 + \log n$  & $\checkmark$ & $L_2, L_\infty$ \\  %
\midrule
Flamingo~\cite{ma23flamingo} & \begin{tabular}[c]{@{}l@{}}Regular: $\ell +C$ \\ \Com:  $n + C$ \end{tabular} &  \begin{tabular}[c]{@{}l@{}}Regular: $\ell + C$ \\ \Com:  $n+C$ \end{tabular} & $\ell n  + C n$ & $\ell n +  C n$ & 3 & $\times$ & N/A  \\  %
\midrule
\sys{} (this work) &  \begin{tabular}[c]{@{}l@{}}Regular: $\ell+C$ \\ \Com:  $n + C$\end{tabular}   & \begin{tabular}[c]{@{}l@{}}Regular: $\ell+C $\\ \Com: $n + C $ \end{tabular} & $\ell n + Cn$ & $\ell n + Cn$  &  3 & $\checkmark$ & $L_2, L_\infty$  \\  %
\bottomrule

\end{tabular}
}

\caption[Asymptotic communication and computation cost for one aggregation]{Asymptotic communication and computation cost for one aggregation for
input vector length $\ell$ and $n$ clients. For simplicity, we omit the
asymptotic notation $O(\cdot)$ in the table, and omit security parameter when it
is an additive factor.
The round complexity excludes any setup that is one-time. 
The round complexity of ACORN-detect is counted using the fixed version (Appendix~\ref{s:fix-acorn}).
For the protocols using the ideas of sub-sampling clients (``decryptors''), 
we denote the number of sampled clients as $C$ where $C=o(n)$. 
In Flamingo, \com has asymptotic cost slightly larger than $n$ when dropouts happen. 
Eiffel has a different communication model from all the other works: it assumes a public bulletin board and all clients in the protocol post messages on the board.
Note that this table shows only asymptotics; protocols with matching asymptotics can exhibit markedly different concrete performance depending on the specifics of their cryptographic design. 
}
\label{fig:comparison}
\end{figure*}

In this section, we formally describe our problem setting. A training process
consists of $T$ iterations, running between the server and in total $N$ clients.
Each iteration has the same procedure: $n$ clients (indexed from 1 to $n$) are
selected from the $N$ clients\footnote{The strategy for selecting clients per
iteration depends on the training design and is orthogonal to the secure
aggregation problem (details in \S\ref{s:security}). Typically, a random subset
of clients are randomly selected in each iteration.}, where client $i$ holds
vector $\bdx_{i}$, and the goal is to let the server only learns the sum
$\sum_{i=1}^{n} \bdx_{i}$ and what can be implied by the sum. 

In practice, a sum of all the $n$ clients is hard to guarantee as some clients
(even if they are honest) may stop responding to the server during protocol
execution (e.g., due to power failure or unstable connection); this is usually
referred as {\em passive dropouts}~\cite{bonawitz17practical}. The server must
continue without waiting for them to return; otherwise, the training might be
blocked for an unacceptable duration. Also, a malicious client can {\em actively
deviate} from the protocol (e.g., sending incorrect messages, using invalid
inputs). Therefore, a more precise goal in this paper is to compute the sum of
the input vectors from the largest possible set of honest and online clients. We
first introduce our setting below and then formally describe the desired
properties.

\paragraph{Communication model.} Clients are heterogeneous devices with varying
reliability (e.g., cellphones, laptops) and may stop responding due to device or
network failures. We assume an implicit distribution for client response delay.

Each client communicates with the server through a private, authenticated
channel. Private messages between clients are relayed via the server and
encrypted under the recipient's public key; in this work we assume a public key
infrastructure (PKI) as in prior works~\cite{bonawitz17practical, bell20secure,
ma23flamingo}. Public messages are signed with the sender's signing key
(derived from the PKI). 

Our protocol proceeds in {\em rounds}, starting with the server. A round trip
involves the server sending messages to clients, waiting for a fixed time to
collect responses in a message pool, processing them, and proceeding to the next
round.

\paragraph{Threat model.} 
We assume the server is honest-but-curious: it follows the protocol but tries to
  learn individual client's input. We assume the server colludes with up to
  $\eta^*$ fraction of the total $N$ clients, and that the corrupted clients
  (including those in the committee $\comset$ below) may deviate arbitrarily
  from the protocol. Corruption is static: the corrupted \coms are fixed before
  $\comset$ is sampled and for the lifetime of their keys, and the corrupted
  clients of an aggregation before it starts. For each aggregation on $n$
  clients, we assume $\eta$ fraction of them are corrupted, and up to $\delta$
  fraction of may passively drop out (excluding the adversarial clients who on
  purpose drop out). The corruption rate $\eta$ in practice is typically small,
  and prior work in secure aggregation mostly experiments with up to
  10\%~\cite{ma23flamingo, bell22acorn}. The dropout rate depends on the chosen
  waiting time, and can be tuned according to the application requirements. Our
  parameters in \S\ref{s:eval} assume corruption and dropout rates of at most 5\%.

\sys{} requires sub-sampling a set $\comset$ from $N$ clients during setup to
assist secure aggregation. Let $\eta_\comset$ and $\delta_\comset$ denote the
corruption and dropout rates in $\comset$ respectively. The properties of our
aggregation protocol (privacy and robustness) hold when at most a $\delta$
fraction of the $n$ clients drop out and at most an $\eta$ fraction are
corrupted, and condition (v) of Theorem~\ref{thm:main} holds for $\comset$. The
sampling algorithm guarantees the latter condition for $\comset$ holds if
$\eta^*$ remains within certain bounds (Remark~\ref{rmk:election}). 

\subsection{Properties}\label{s:properties}

We formally give the properties that we aim to achieve in
\sys{} for a single aggregation (computing one sum over $n$ clients).  
Discussion for computing multiple sums over different sets of clients is given in Section~\ref{s:security}.

\begin{definition}[Privacy]\label{def:privacy}
  We say an aggregation protocol has privacy against a class of adversaries if
  the protocol realizes the ideal functionality in Figure~\ref{fig:ideal-sum} in
  the presence of any adversary of that class. 
\end{definition}

\begin{figure}[!tb]
        \resizebox{\columnwidth}{!}{\begin{tcolorbox}[colback=white, coltitle=white, colframe=black, enhanced, boxsep=1mm, left= 0mm, right=0.5mm, title={\textbf{\small Functionality $\mathcal{F}$}}]
           \linespread{1.2}
           \small
           
           Parties: A set of $n$ clients $P_1, \ldots, P_n$ and a server $S$ (the \coms have no input or output). 
           
           Notation: Let corruption rate be $\eta$ and dropout rate be $\delta$, both
           among $\mathcal{P} = \{P_1, \ldots, P_n\}$. Let $\Adv$ be the adversary
           corrupting $S$ and the set of clients of size $\eta n$, $\textsf{Cor}$.

           \begin{itemize}
               \item  $\mathcal{F}$ and $\Adv$ receive a set of dropout clients $\mathcal{O}
                       \subset \mathcal{P}$ where $|\mathcal{O}|/|\mathcal{P}|\le
                       \delta$, when their messages arrive (if at all), and when each \com stops responding. $\mathcal{F}$ receives $\bdx_{i}$ of client $P_i
                       \in\mathcal{P}\backslash \mathcal{O}$.
            
               \item   $\mathcal{F}$ asks $\Adv$ for a set $\mathcal{E}$ with the
                       requirements that: $|\mathcal{E} \cup \mathcal{O} | / n \le
                       \delta$. 
                
                \item  If $\Adv$ replies $\mathcal{F}$ with a set $\mathcal{E}$ that satisfies
                       the requirement, then $\mathcal{F}$ outputs $\bold{z} =
                       \sum_{i\in \mathcal{P}
                       \backslash ( \mathcal{E}\cup \mathcal{O} \cup \mathsf{Cor})}
                       \bdx_{i}$ to $\Adv$.     
                       Otherwise, $\mathcal{F}$ send \texttt{terminate} to all
                       parties. 
            \end{itemize}
                      
           \end{tcolorbox}}
       \caption{Ideal functionality for one aggregation.
       We follow the definition in prior works~\cite{bonawitz17practical,
       bell20secure} 
       assuming an oracle gives the dropout set, with these times, to $\mathcal{F}$ and 
       adversary $\Adv$ can also query the oracle; a client drops out if its
       Round 1 message does not reach the server within its waiting time. }
       \label{fig:ideal-sum}
       \end{figure}

\begin{definition}[Dropout resilience]\label{def:dropout-resilience}
  We say an aggregation protocol on $n$ clients with inputs $\bdx_1, \ldots, \bdx_n$ has dropout resilience if, when all clients follow the protocol and a set $\mathcal{X} \subseteq[n]$ of clients remains online throughout the aggregation,
  the server should output $\sum_{i\in \mathcal{I}} \bdx_i$ where $\mathcal{X} \subseteq \mathcal{I} \subseteq [n]$. 
  
\end{definition}

The salient aspect of \sys{} is disruption
resistance: the coalition of malicious clients can affect the aggregation
result only by misreporting their private inputs. 
This is formalized in Definition~\ref{def:robustness} and~\ref{def:disruption-resistance} below.
A pre-condition for disruption resistance is that this protocol can be completed regardless how adversarial clients act.

\begin{definition}[Robustness]\label{def:robustness}
  Let $f$ be an aggregation function that takes in $n$ inputs $\bdx_1,
      \ldots, \bdx_n$.
      Let $\mathcal{Z}$ be the set of all possible inputs for $f$.
      We say that an $n$-client aggregation protocol has {\em robustness} if, 
      when the server follows the protocol, for every number of $m$ adversarial clients (with
      $0\le m \le \eta n$, at most $\eta_\comset C$ adversarial \coms, and at
      most $\delta n$ clients dropping out) and for every choice of honest
      clients' inputs in $\mathcal{Z}$, the protocol always outputs to the server
      a value in the set $\{ f(\mathcal{I}_{\mathrm{honest}}, \mathcal{I}_{\mathrm{adv}}) \ |
      \ \mathcal{I}_{\mathrm{adv}}\in \mathcal{Z}^{m}\}$, where
      $\mathcal{I}_{\mathrm{honest}}$ are the inputs of the honest clients that
      do not drop out.
      
\end{definition}

  \begin{definition}[Disruption resistance~\cite{corrigan-gibbs17prio}]
  \label{def:disruption-resistance}
      Let $f$ be an aggregation function that takes in $n$ inputs $\bdx_1,
      \ldots, \bdx_n$.
      We say a protocol has {\em disruption resistance} if Definition~\ref{def:robustness}
      holds when $\mathcal{Z}$ is restricted to a user-defined set of valid inputs
      ($\bdx$ is a valid input for $f$ if $\bdx\in \mathcal{Z}$).
  \end{definition}

  In this work, we consider the aggregation function $f$ to be a simple sum
  function $f(\bdx_1, \ldots, \bdx_n) = \bdx_1 + \cdots + \bdx_n$, and $\mathcal{Z}$
  to be all possible vectors with application-defined $L_2, L_\infty$ norms.

\section{Preliminaries}\label{s:prelim}

\paragraph{Notation.}
Let $[z]$ denote the set $\{1, 2, \ldots, z\}$. We use $[a, b]$ to denote the
set $\{x\in\mathbb{N}: a\le x \le b\}$. We use bold lowercase letters (e.g.
$\bdu$) to denote vectors and bold upper case letters (e.g., $\bdA$) to denote
matrices. Unless specified, vectors are column vectors. Given a value $\alpha$
and a vector $\bdv$, we use $\alpha \bdv$ to denote multiplying $\alpha$ to
every coordinate of $\bdv$. For distribution $\calD$, we use $a \leftarrow
\calD$ to denote sampling $a$ from $\calD$.  
For a vector $\bdv$, we use $\lfloor \bdv \rceil_{c}$ to denote rounding each entry of $\bdv$ to nearest multiples of $c$. 
For two vectors $\bdv_1$ of length $\ell_1$, $\bdv_2$ of length $\ell_2$, we use $\bdv_1 | \bdv_2$ to denote the concatenation of them which is a vector of length $\ell_1 + \ell_2$.
We use $\|\bdv\|_2$ to denote $L_2$ norm of $\bdv$ and use $\|\bdv\|_\infty$ to denote the largest entry in $\bdv$. 
We use $\mathbb{F}$ to denote a field.

\paragraph{Regev's encryption.}
Our construction utilizes the key-and-message homomorphism Regev's
encryption~\cite{regev05lattices}; we give the details below. The Regev's scheme
is parameterized by a security parameter $\lambda$, a plaintext modulus $p$, and
a ciphertext modulus $q$, and number of LWE samples $m$. Given a secret key $\bds \xleftarrow{\$}
\Z_q^{\lambda}$, the encryption of a vector $\bdx \in \Z_p^m$ is 
\[ (\bdA, \bdc) := (\bdA, \bdA \bds + \bde + \Delta \cdot \bdx), \] where $\bdA
\xleftarrow{\$} \Z_q^{m\times \lambda}$ is a random matrix ($m>\lambda$), $ \bde
\xleftarrow{\$} \chi^m$ is an error vector and $\chi$ is a discrete Gaussian
distribution, and $\Delta := \floor{q/p}$. Decryption is computed as $(\bdc -
\bdA \bds) \bmod q$ and rounding each entry to the nearest multiples of
$\Delta$, and then divide the rounding result by $\Delta$. The decrypted result
is correct if entries in $\bde$ are less than $\Delta/2$.

\paragraph{Packed secret sharing.}
In standard Shamir secret sharing~\cite{shamir1979share}, a secret $\secret \in \mathbb{F}$ is hidden 
as the constant term of a polynomial $\polynomial(x) = a_0 + a_1 x + \cdots + a_d x^d$ where $a_0 = \secret$
and $a_1,\ldots, a_d$ are randomly sampled from $\mathbb{F}$. Given $n$ parties, the share for
party $i \in [n]$ is $\polynomial(i)$, and any subset of at least $d+1$ parties can
reconstruct $\secret$ and any subset of $d$ shares are independently random. 

In packed secret sharing~\cite{STOC:FraYun92}, one can hide multiple secrets using a
single polynomial. Specifically, let $\mathbb{F}$ be a field of size at least
$2n$, let $k$ be the number of secrets packed in one sharing, and let $\tau$ be
the privacy threshold. Packed Shamir
secret sharing of $(v_1, \ldots, v_k) \in \mathbb{F}^k$ first chooses a random
polynomial $\polynomial(\cdot) \in \mathbb{F}[X]$ of degree at most $\tau+k-1$
subject to $\polynomial(0)=v_1, \ldots, \polynomial(-k+1)=v_k$, and then sets
the share $\share_i$ for party $i$ to be $\share_i=\polynomial(i)$ for all
$i\in[n]$. Reconstruction of a degree-$(\tau+k-1)$ sharing requires at least
$\tau+k$ shares from $\share_1, \ldots, \share_n$. Note that the corruption
threshold is now $\tau$ even if the degree is $\tau+k-1$, i.e., any $\tau$
shares are independently random, but any $\tau+1$ shares are not.

\paragraph{Shamir sharing testing.}
Looking ahead, we will also use a probabilistic test for Shamir's secret shares,
called SCRAPE test~\cite{ACNS:CasDav17}. To check if $(\share_1, \ldots, \share_n)\in
\mathbb{F}^n$ is a Shamir sharing over $\mathbb{F}$ of degree $d$ (namely there
exists a polynomial $p$ of degree $\le d$ such that $p(i) = \share_i$ for $i = 1,
\ldots, n$), one can sample $w_1, \ldots, w_n$ uniformly from the dual code to
the Reed-Solomon code and check if $w_1 \share_1 + \cdots + w_n \share_n = 0$ in
$\mathbb{F}$.  

To elaborate, let $c_i:=\prod_{j\in[n]\setminus \{i\} }(i-j)^{-1}$ and
$m^\ast(X):=\sum_{i=0}^{n-d-2} m_i\cdot X^i\gets_\$ \mathbb{F}[X]_{\leq n-d-2}$
(a random polynomial over $\mathbb{F}$ of degree at most $n-d-2$). Now, let
$\bfw:=(c_1\cdot m^\ast (1),\ldots,c_n\cdot m^\ast(n))$ and
$\bm{\share}:=(\share_1,\ldots,\share_n)$. Then, 
\begin{myitemize2}
    \item If there exists $ \polynomial \in\mathbb{F}[X]_{\leq d}$ such that
    $\share_i=\polynomial(i)$ for all $i\in[n]$, then $\langle
    \bfw,\bm{\share}\rangle=0$. 
    \item Otherwise, $\Pr[\langle \bfw,\bm{\share} \rangle=0]=1/|\mathbb{F}|. $
\end{myitemize2} 
In other words, if $(\share_1, \ldots, \share_n)$ is not a Shamir sharing of degree $d$
then the test will only pass with probability $1/|\mathbb{F}|$.

\paragraph{Pedersen and vector commitment.}
Let $\G$ be a group of order $q$, and $G, H$ be two generators in $\G$. A
Pedersen commitment to a value $v\in\Z_q$ is computed as $\commit_{G}(v):=  G^v
H^r$, where the commitment randomness $r$ is uniformly chosen from $\Z_q$. We
use $\commit_G(\cdot)$ notation because later in our protocol we compute
commitments with different generators.

We can also commit to a vector $\bdv=(v_1, \ldots, v_L) \in \Z_q^L$ as follows:
let $\bold{G}= (G_1, \ldots , G_L)$ be a list of $L$ random generators in $\G$,
define $\commit_{\bold{G}}(\bdv) :=  G_1^{v_1}\cdots G_L^{v_L} \cdot H^r$, where
$r$ is randomly chosen from $\Z_q$; our notation $\commit_{\bold{G}}(\cdot)$
implictly assumes a public $H$ and a private $r$ are included.

\paragraph{Inner-product proof.}
The inner product argument is an efficient proof system for the following
relation: given two vector commitments $\commit(\bda), \commit(\bdb)$ known to
both a prover and a verifier and a public value $c$, the prover can convince the
verifier that $\langle \bda, \bdb \rangle = c$.
Bulletproof~\cite{bunz18bulletproof} gives non-interactive inner-product proof
using Fiat-Shamir, with proof size $O(\log L)$ and prover/verifier cost $O(L)$.
In our protocol, every Fiat-Shamir challenge in a client's proofs also hashes
the public parameters of the training iteration (the seed of $\bdA$) and the
client's identity, so a proof cannot be reused in another iteration or by
another client.

For ease of presentation later, we introduce the following notations for proof.
A proof system $\Pi$ consists of a tuple of algorithms $(\prover, \verifier)$
run between a prover and verifier. An argument to prove can be described with
public inputs/outputs $\zkio$, a statement to be proved $\zkstate$, and a private
witness $\witness$. Given a proof system $\Pi$, the prover can generate a proof
$\pi \leftarrow \Pi.\prover(\zkio, \zkstate, \witness)$ and the verifier checks
the proof by $b\leftarrow \Pi.\verifier(\zkio, \zkstate, \pi)$ where $b \in
\{0,1\}$ indicates rejecting or accepting $\pi$. For example, for proving inner
product of $\bda$ and $\bdb$, we set the constraint system to be
 \[
 \{
 \zkio: (\commit(\bda), \commit(\bdb), c), \ 
 \zkstate: \langle \bda, \bdb \rangle=c , \ 
 \witness : (\bda, \bdb) 
 \}.
\]
Denote the inner product proof system as $\Pi_{\text{ip}}$, the prover runs $\pi
\leftarrow \Pi_{\text{ip}}.\prover(\zkio, \zkstate, \witness)$ and the verifier
runs $b\leftarrow \Pi_{\text{ip}}.\verifier(\zkio, \zkstate, \pi)$. The
algorithms $\Pi_{\text{ip}}.\prover$ and $\Pi_{\text{ip}}.\verifier$ both have
complexity linear to the length of $\bda$ (or $\bdb$) and $\pi$ has logarithmic
length of $\bda$ (or $\bdb$). We will also prove {\em linear-relation}, and we
denote the proof system as $\Pi_{\text{linear}}$ and the constraint system will
be
\[
 \{
 \zkio: ( \commit(\bdb), \bda, c), \
 \zkstate: \langle \bda, \bdb \rangle=c , \ 
 \witness : \bdb 
 \}. 
\]
To differentiate the two proof systems, we call the former (that needs to
commit to both vectors in the inner product) as quadratic proof and the later
(that only needs to commit to one vector in the inner product) as linear proof.

\section{Protocol design}\label{s:overview}

Now we describe our construction for computing a single sum (one
iteration in the training). 
Our full protocol is given in Figures~\ref{fig:full} and
\ref{fig:full2}; below we describe our main technical
ideas.

\subsection{A two-layer secure aggregation}\label{s:2layer} 

The key idea is to reduce an aggregation for long vectors to an aggregation for
short vectors. 
To substantiate this idea, we utilize the key-and-message homomorphism of
Regev's encryption. 

Given two Regev ciphertexts $(\bdA, \bdc_1) , (\bdA, \bdc_2)$ of vectors $\bdx_1,
\bdx_2$ under the key $\bds_1, \bds_2$ with noise $\bde_1, \bde_2$, the tuple
$(\bdA,  \bdc_1 + \bdc_2)$ is an encryption of $\bdx_1 + \bdx_2$ under the key
$\bds_1+\bds_2$. 
The ciphertext $(\bdA, \bdc_1 + \bdc_2)$ can be properly decrypted if
$\bde_1 + \bde_2$ is small. 
Note that computing $\bdc_1 + \bdc_2$ is very efficient---it is simply vector
addition.

For ease of presentation later, we define a tuple of algorithms $(\Enc, \Dec)$
parameterized by $(p,q,\lambda,m,\bdA\in\Z_q^{m\times \lambda})$ as follows: 
\begin{myitemize2}
    \item $\Enc(\lwesecret, \bdx) \rightarrow \bdy$: on input a secret key
    $\lwesecret \in \Z_q^\lambda$ and a message $\bdx\in\Z_p^m$, output $\bdy :=
    \bdA \cdot \bds + \bde + \Delta \cdot \bdx$, where $\Delta=\floor{q/p}$. 

    \item $\Dec(\lwesecret, \bdy) \rightarrow \bdx'$: on input a secret key
    $\lwesecret \in \Z_q^\lambda$ and a ciphertext $\bdy \in\Z_q^m$, output
    $\bdx' := \lfloor \bdy-\bdA \bds \rceil_{\Delta}$.
\end{myitemize2} 

Now we describe how our protocol work at a high level. Each client $i \in [n]$
 holds an input vector $\bdx_i \in \Z_p^\ell$, and w.l.o.g. assume $\ell$ equals
 the LWE parameter $m$.\footnote{One should first fix $\ell$
 required in applications, and choose the set of LWE parameters to get a target
 security level.} The client samples a Regev encryption key $\lwesecret_i \in
 \Z_q^\lambda$ and sends the encrypted vector $\bdy_i = \Enc(\lwesecret_i,
 \bdx_i):=\bdA \lwesecret_i+\bde_i+\Delta\cdot \bdx_i$ to the server. Note that
 $\lambda \ll \ell$. The server computes the sum of $\bdy_i$'s as 
\[
\bdy := \sum_{i \in [n]} \bdy_i = \sum_{i \in [n]} \bdA \lwesecret_i+\bde_i+\Delta\cdot \bdx_i = \bdA \sum_{i \in [n]} \lwesecret_i+ \Delta \sum_{i \in [n]} \bdx_i+\sum_{i \in [n]} \bde_i,
\]
To reconstruct $\sum_{i \in [n]} \bdx_i$, the server needs $\lwesecret :=
\sum_{i=1}^n \lwesecret_i$ to decrypt $\bdy$,  and the decryption succeeds if
$\|\sum_{i\in[n]} \bde_i\|_\infty < \Delta/2$. We call the sum of Regev's ciphertexts
$\bdy_i$'s as {\em outer aggregation}, and next we discuss {\em inner
aggregation} where the server gets sum of $\lwesecret_i$'s. 

The inner aggregation could be instantiated with a naive secure multi-party
computation over the $n$ clients: each client $i$ secret shares
$\lwesecret_i$ coordinate-wise to all the other clients (the shares are
encrypted using public keys of the recipient clients and sent through the
server),
and each client adds up the shares which is then sent to the server for
reconstruction of $\lwesecret$.
However, this naive approach has per client communication $O(n \lambda)$ and server communication
$O(n^2 \lambda)$. The former is too much for a client given $\lambda$ is a
security parameter and is typically from a few hundreds to a thousand; and the latter
is too much for the server because it is quadratic in $n$. 

We reduce the communication complexity with two techniques: 1) let $C\ll n$, sample
$C$ clients from the whole population as {\em \coms} to assist with unmasking the aggregation result\footnote{We show that $C$ can be 
polylogarithmic in $n$ to have this work (\S\ref{s:security}).}; 2) each client $i$
uses packed secret sharing (\S\ref{s:prelim}) to share its secret vector
$\lwesecret_i$ to the \coms, so that each \com receives one share from a client. 
The combination of these two techniques result in 
communication complexity per regular client $O(C)$ and per \com $O(n)$, 
and the server communication complexity is $O(C n)$.
Although the \com cost is linear in $n$, we show in Section~\ref{s:eval} that 
the cost is modest for the $n$ needed in federated learning.

This inner-outer paradigm has a key advantage in handling dropouts, unlike the
pairwise masking approach used in prior works~\cite{bonawitz17practical,
bell20secure, ma23flamingo} which incurs extra rounds. Specifically, if a
client drops out during the outer aggregation (for sending $\bdy_i$), the
server can safely ignore the client without affecting subsequent steps. If a
\com client drops out during the inner aggregation, the server can still
reconstruct $\lwesecret$ due to the threshold nature of secret sharing.

In the next few sections, we discuss how to make this simple protocol robust 
against adversarial clients (which may include some of the \coms) in two parts: 1) proof of
linear computation in outer aggregation and 2) an agreement protocol for inner aggregation. 

\begin{remark}
As observed in a few works in orthogonal areas~\cite{henzinger23one,
davidson23frodo}, Regev's encryption remains secure even if $\bdA$ is made
public and the same matrix $\bdA$ is used to encrypt polynomially many messages,
as long as the secret key $\bds$ and the noise $\bde$ are independently chosen
in each instance of encryption. Therefore for our protocol, $\bdA$ can be
generated by a trusted setup~\cite{das22spurt, cloudflare-beacon} (to reduce the
cost of randomness generation, the trusted entity can generate a short random
string and expand it to $\bdA$ using pseudorandom generator). This only runs
once: $\bdA$ for iteration $t$ is expanded from this string and $t$, so an attack
on $\bdA$ cannot be amortized over the keys of all iterations. Also, a client $i$ can pre-compute $\bdA\bds_i$
prior to knowing their input $\bdx_i$.
\end{remark}

\begin{remark}[Selecting \coms]\label{rmk:election}
So far, we assume the existence of a set $\comset$ of \coms with corruption rate
 $\eta_\comset$ before the aggregation starts. In fact, $\eta_\comset$ depends
 on the exogenous parameter $\eta^*$ and how we sample $\comset$. One can use
 the sampling approach in Flamingo~\cite{ma23flamingo} which assumes there is a
 random beacon.  
Alternatively, we can also use 
a sampling protocol by Alon et al.~\cite{GMPC22} which does not
 require this assumption. 
Readers can refer to the two papers~\cite{ma23flamingo, GMPC22} for formal
analysis of the relation between $\eta_\comset$ and
 $\eta^*$.  
\end{remark}

\subsection{Proof of client computation}\label{s:robust} 

Our high-level idea is ``commit-and-proof'': each client sends commitments to
its private values (e.g., commitment to $\bds_i$) together with a proof of the
following relations. Let $\bold{F}, \bold{G}, \bold{H}$ be vectors of group
generators in $\G$ of length $\lambda, \ell, \ell$ respectively. Suppose client
$i$ sends to the server $\commit_{\bold{F}}(\lwesecret_i),
\commit_{\bold{G}}(\bde_i), \commit_{\bold{H}}(\bdx_i)$, in addition to $\bdy_i$
as specified in the outer aggregation. (Section~\ref{s:proof-norms} refines
this for the LWE error: only its high digit $\bde_i^{\mathrm{hi}}$ is committed,
and the encryption relation is proven together with the bound on the low
digit.) The client proves to the server that:
\begin{myenumerate2}
    \item For the outer aggregation, $\bdy_i := \bdA \cdot \lwesecret_i + \bde_i
    + \Delta \cdot \bdx_i \bmod q$, with $\bde_i$ having small $L_\infty$ norm.
    \item For the inner aggregation, the client secret-shares $\lwesecret_i$ to
    the \coms using a polynomial of degree $d$ (the degree $d$ is fixed by the
    threat model parameters, see \S\ref{s:security}).
\end{myenumerate2}

Next, we express these requirements (except the norm condition which is
non-linear) as inner-product relations. We will address proving the norms in
Section~\ref{s:proof-norms}. We set LWE modulus $q$ to match the field size of
the commit-and-proof system.

\paragraph{Proving the first statement.}
At the first glance, we need to prove that each coordinate of $\bdy_i$ equals
the corresponding coordinate of the RHS computation result. This would require
$\ell$ proofs, one for each coordinate. We instead use a polynomial checking
technique from Schwartz-Zippel Lemma to compress the proofs to a single one. In
particular, if we want to check if two vectors of length $\ell$ over $\Z_q$ are
equal, we view each vector (e.g., $\bdy_i$) as coefficients of a degree-$(\ell-1)$
polynomial and check if the evaluation of the two polynomials on a random point
are equal. If they are indeed not equal, then the evaluation will be different
except probability $\ell/q$. Formally, let $r \in \Z_q$ be a random challenge
value picked by the server (who is the verifier), and let $\bdr=(r^0, r^1,
\ldots, r^{\ell-1})$. Let $c = \langle \bdy_i, \bdr \rangle$, and $c$ is a
public value since $\bdy_i$ and $\bdr$ are both public (known to both the client
and the server). If the client can prove to the server that 
\[ c = \langle \bdA^\top \bdr \ |\ \bdr \ |\ \Delta \bdr, \ \lwesecret_i \ |\
\bde_i \ |\ \bdx_i \rangle \text{ in }\Z_q , \] then the server will be
convinced that $\bdy_i := \bdA \cdot \lwesecret_i + \bde_i+ \Delta \cdot
\bdx_i$, and there will only be $\ell/q$ probability that the client is
dishonest but the server is convinced. The inner product argument comes from the
following: 
$$
\begin{aligned}
      \langle \bdy_i, \bdr \rangle 
    = & \langle \bdA \lwesecret_i, \bdr \rangle + \langle \bde_i, \bdr \rangle + \Delta \langle \bdx_i, \bdr \rangle \\
    = &\langle \bdA^\top \bdr, \lwesecret_i \rangle + \langle  \bdr , \bde_i \rangle + \langle  \Delta \bdr, \bdx_i \rangle \\
    = & \langle \bdA^\top \bdr \ | \ \bdr \ |\ \Delta \bdr, \ \lwesecret_i \ |\ \bde_i \ |\ \bdx_i \rangle. 
\end{aligned}
$$
Also, note that $\bdA^\top \bdr \ |\ \bdr \ |\ \Delta \bdr$ is public, the
client only needs to do a linear proof, where the witness is under the
commitment $\commit_{\bold{F} | \bold{G} | \bold{H}} (\lwesecret_i | \bde_i |
\bdx_i)$. 

\paragraph{Proving the second statement.}
Recall that the client sends $\commit_{\bold{F}}(\lwesecret_i)$ to the server in
the outer aggregation, and now we want to ensure that the shares that the
\coms received (for the inner aggregation) are indeed the Shamir shares of
this committed $\lwesecret_i$. Here we use the SCRAPE test
(\S\ref{s:prelim}) to express this constraint as an inner product relation; this
test seamlessly works with packed secret sharing.  

Formally, suppose client $i$ has a packed Shamir sharing of $\lwesecret_i$ as a
vector of length $C$ (recall that there are $C$ \coms)
 \[\bm{\share}_i = (\share_i^{(1)}, \ldots, \share_i^{(C)}),\] which the client
claims is a sharing of degree $d$ over $\Z_q$. We observe that checking if
$\bm{\share}_i$ is a packed sharing of $\lwesecret_i$ is equivalent to checking
if $( \bm{\share}_i \ | \ \lwesecret_i)$ is a sharing of length $C+\lambda$ of a
degree-$d$ polynomial. Therefore, we let the client commit to $\bm{\share}_i$
under a new generator vector $\bold{K}=(K_1, \ldots, K_C) \in \G^C$, and sends
$\commit_{\bold{K}}(\bm{\share}_i)$ to the server in the outer aggregation as
well. Then the client invokes a linear-relation proof that 
\[\langle \bm{\share}_i \ | \ \lwesecret_i , \ \bfw \rangle =0 \text { in }
\Z_q,\] where $\bfw:=(w^{(1)}, \ldots, w^{(C+\lambda)})$ is sampled uniformly
random from some code space (details in \S\ref{s:prelim}) and is public (known
to both the server and client). In our setting, we cannot let the client choose
$\bfw$ since the client may be malicious, so we apply the Fiat-Shamir transform
and have client $i$ derive $\bfw$ by hashing $\commit_{\bold{F}}
(\lwesecret_i)$ and each of the commitments to the shares below.

Up to this point, we have not guaranteed the shares received by the \coms are
consistent with the commitment $\commit_{\bold{K}}(\bm{\share}_i)$. The reason
is that the client could in fact send to a \com a share (under the encryption)
that is different from what was committed to. Therefore, instead of having the
client send $\commit_{\bold{K}}(\bm{\share}_i)$ to the server, we have the
client send commitments to each coordinate of $\bm{\share}_i$, namely
$\commit_{K_1}(\share_i^{(1)}), \ldots,\commit_{K_C}(\share_i^{(C)}) $.
Each $\commit_{K_j}(\share_i^{(j)})$ has its own randomness $r_i^{(j)}$
(\S\ref{s:prelim}), which the client sends to the $j$-th \com together with
$\share_i^{(j)}$. The server can still verify the proof for packed sharing,
but not on the plain product of the $C$ commitments: keeping it fixed, a client
could shift the shares of the \coms that check their commitments and cancel
the shifts in the commitment of a \com that never checks its own (corrupt or
offline), e.g., by a polynomial of degree $d$, so that the sums decode to a
wrong $\lwesecret$. As in \S\ref{s:proof-norms}, the proof instead runs on
$\commit_i^{\bm{\gamma}} := \commit_{\bold{F}}(\lwesecret_i)^{\gamma_0}\prod_{j}
\commit_{K_j}(\share_i^{(j)})^{\gamma_j}$ for Fiat-Shamir challenges
$\bm{\gamma}$ derived after all commitments are fixed, with the matching entries
of $\bfw$ divided by $\gamma_j$; by Schwartz-Zippel, it binds each
$\commit_{K_j}(\share_i^{(j)})$ to the $j$-th share of a degree-$d$ sharing of
$\lwesecret_i$.

For those clients whose proofs are valid (for both $\Enc$ computation and SCRAPE
test), the server forwards their commitments to the corresponding \coms where
$\share_i^{(j)}$ is intended for the $j$-th \com. Then each \com $j$
verifies if the received share and randomness (after decryption) are
consistent with the commitment $\commit_{K_j}(\share_i^{(j)})$.

\subsection{Agreement protocol for \coms}\label{s:agreement}

So far, each \com can identify a set of \emph{valid} shares, which are
consistent with the vector commitments. However, this alone is not enough for
the server to obtain the correct sum, as illustrated in the following example.
Suppose there are three clients \(P_1\), \(P_2\), and \(P_3\), with \(P_3\)
being malicious, and three \coms \(H_1\), \(H_2\), and \(H_3\), with \(H_3\)
being malicious. Each client shares a secret using 2-out-of-3 Shamir sharing. If
\(P_3\) sends a valid share to \(H_1\) but an invalid share to \(H_2\), the
\coms will form sets of clients with valid shares as:

\begin{itemize}\itemsep0em
    \item \(H_1\) forms \(\{P_1, P_2, P_3\}\),
    \item \(H_2\) forms \(\{P_1, P_2\}\),
    \item \(H_3\) can form any arbitrary set, e.g., \(\{P_3\}\).
\end{itemize}
If each \com adds up the valid shares locally, the resulting values will not
reconstruct the sum of the secrets from any set of clients. The server can only
reconstruct a meaningful sum if the \coms sum shares from the \emph{same set}
of clients, i.e., the sum of the secrets from that specific set.

Our goal is to ensure that all honest \coms agree on the same set of clients
with valid shares. Although this seems like a consensus problem that may
require many rounds, we can utilize the server as the central coordinator where 
any \com can complain to the server about an invalid share from a malicious client. 
However, the server does not know
whether this complaint is fake as it could come from a malicious \com. 
To help the server distinguish between valid complaints and fake complaints, 
we use a simple proof of decryption to make complaints {\em verifiable}. 
This is inspired by Benhamouda et al.~\cite{BHKMR24}.

We let the clients in the first round send their shares to \coms encrypted
under a public-key encryption scheme $\AsymEnc$ with labels: client $i$ sends
to \com $j$, via the server, the ciphertext $\mathsf{ct}_{i,j} =
\AsymEnc(PK_j, (t,i,j), (\share_{i}^{(j)}, r_i^{(j)}))$ of its share and of
the randomness of its share commitment, under the label $(t,i,j)$, where $t$
is the ID of the aggregation (unique under the \coms' keys,
Figure~\ref{fig:full}) and $PK_j$ is $j$'s public key. If $\mathsf{ct}_{i,j}$ is not
a valid ciphertext under this label, or its decryption is not consistent with
the commitment $\commit_{K_j}(\share_i^{(j)})$, then \com $j$ sends a {\em
verifiable complaint} to the server. For an invalid ciphertext, the ciphertext
itself is the evidence, since validity is publicly checkable. For a valid one,
the complaint consists of the decryption $(\share_{i}^{(j)}, r_i^{(j)})$ and a
zero-knowledge proof that it is the decryption of $\mathsf{ct}_{i,j}$ under
the secret key $SK_j$ of $PK_j$. The server knows $\mathsf{ct}_{i,j}$ and the
commitment from the first round, so it can check that the ciphertext is
invalid, or that the proof verifies and the decryption is inconsistent with
the commitment. The server only accepts complaints that pass these checks, and
then informs the \coms of the lying clients to remove them from their sets. At
this point, all the honest \coms agree on a same set. Now they can add up the
shares of this set.
Finally the server can perform error correction on the Shamir shares from the
\coms, and reconstruct the sum even if there are bogus shares from malicious
\coms.

A complaint must not turn a \com into a decryption oracle. If a complaint
revealed the decryption of whatever ciphertext it is about, a client colluding
with the server could submit an honest client's ciphertexts (or mauled
versions of them) as its own, and read the honest client's shares, and with
enough of them its input, off the honest \coms' complaints. This is why
$\AsymEnc$ has labels, publicly verifiable ciphertexts, and security against
chosen-ciphertext attacks: under the colluding client's label, a copied or
mauled ciphertext is invalid and its complaint reveals nothing, and the
decryption of a valid ciphertext under label $(t,i,j)$ tells nothing about
ciphertexts under other labels.

We instantiate $\AsymEnc$ with TDH2~\cite{EC:ShoGen98} (with a single
decryptor) over the group $\G$ of the commit-and-prove system; it is IND-CCA
secure with labels in the random-oracle model under the DDH assumption. Let
$g, \bar g$ be generators of $\G$ with unknown discrete-log relation ($\bar g$
is obtained by hashing into $\G$, modeled as a random oracle), and let
$SK_j = x_j$ and $PK_j = h_j = g^{x_j}$. A message $m$ is encrypted under label
$L$ as $(c, u, \bar u, e, f)$, where $u = g^k$ and $\bar u = \bar g^k$ for a
random $k$, $c = m \oplus \mathcal{H}_1(h_j^k, u)$, and $(e, f)$ is a
Fiat-Shamir proof that $\log_g u = \log_{\bar g} \bar u$ whose challenge also
hashes $c$ and $L$; the ciphertext is valid if this proof verifies. To
complain about a valid ciphertext, a \com reveals $K = u^{x_j}$ with a
Chaum-Pedersen proof~\cite{C:ChaPed92} that $\log_g h_j = \log_u K$, and the
server computes $m = c \oplus \mathcal{H}_1(K, u)$ and checks that it is
inconsistent with the share commitment. As $K$ is determined by $h_j$ and $u$,
a corrupt \com cannot make an honest client's ciphertext decrypt to anything
else; registering $PK_j$ takes a proof of knowledge of $x_j$. With a share and
its commitment randomness as the 64-byte message, a ciphertext takes 192
bytes, and a complaint about a valid ciphertext 96 bytes.

\subsection{Integrating with proof of norms}\label{s:proof-norms}

Our protocol can be seamlessly integrated with existing techniques on proof of
norms~\cite{bell22acorn, GHL22}. Now we describe how a client proves a vector
$\bdx_i$ has bounded $L_2, L_\infty$ norms; for the LWE error vector $\bde_i$
that we mentioned earlier, we use a two-digit variant at the end of this section. Below, we represent
an integer as a number in $\Z_q$ where the negative numbers are in $(-q/2, -1]$
and the positive numbers are in $[0, q/2]$. We will use the term ``ZKPoK'' to
denote zero-knowledge proof of knowledge.

We now present a primitive that we will use as blackbox called {\em
approximate proof of $L_\infty$-smallness}~\cite{lyubashevsky2021shorter}. With
internal bound $B_{\mathrm{apx}}$, it is complete for $\|\bda\|_\infty <
B_{\mathrm{apx}}$ and sound against $\|\bda\|_\infty > \gamma
B_{\mathrm{apx}}$; below we choose $B_{\mathrm{apx}}$ so $\gamma
B_{\mathrm{apx}}$ is below the needed wraparound threshold. We defer the
construction to \appref{app:approx}; in short, an approximate proof for a
length-$\ell$ vector invokes a linear proof of length $\ell+\sigma$ where
$\sigma$ is a security parameter; we take $\sigma = 128$ as in~\cite{GHL22}. 

The idea underlying proof of $L_2$ and $L_\infty$ norms is the same: Assuming
the bound $B$ we want to prove is much smaller than $\sqrt{q}$, to prove a
number $x$ is bounded by $B$ in a field $\Z_q$, we find four integers $a_1,
a_2, a_3, a_4$ such that\footnote{Any positive integer $a$ can be decomposed
into four squares and
finding the four squares can be done in $O(\log^2 a)$
time~\cite{rabinshallit}.}
\[ B^2 = x^2 + a_1^2 + a_2^2 + a_3^2 + a_4^2. \]
The proof uses the underlying commit-and-prove systems to show this equality
modulo $q$, and approximate proof to ensure $x, a_1,\ldots, a_4$ are small
enough to avoid wraparound modulo $q$.
For wraparound, it suffices to set the approximate-proof parameters so that
the soundness bound guarantees $x, a_1, \ldots, a_4$ are all smaller than
$\sqrt{q/10}$; then $x^2+ a_1^2 +\cdots + a_4^2 < q/2$.

To prove a length-$\ell$ vector $\bdx$ has $L_2$ norm bounded by $B$, we find
four integers $a_1, \ldots, a_4$ such that  
\[ B^2 = \|\bdx \|_2^2 + a_1^2 + \cdots + a_4^2,\]
and define $\bdv = \bdx | (a_1, \ldots, a_4)$; the prover commits to $(a_1,
\ldots, a_4)$ on generators disjoint from those of $\bdx$, and $\commit(\bdv)$ is
assembled homomorphically from $\commit(\bdx)$ and this commitment. Whenever the
commitment of a linear proof (including the one in the approximate proof) is
assembled homomorphically, directly or through $\commit(\bdv)$ or
$\commit(\vec{z})$ below, from commitments the prover has published, each
published commitment (e.g., $\commit(\bdx)$ and the commitment to $(a_1, \ldots,
a_4)$ separately) is first raised to its own random challenge, derived by
Fiat-Shamir after all of them are fixed, and the matching part of the public
vector is divided by that challenge; by Schwartz-Zippel, the proof then binds
each of these commitments to its own part of the vector, not only their product
(up to components on generators where the public vector is zero, which no
relation reads).
The prover provides a ZKPoK that $\|\bdv\|_2^2 = B^2 \bmod q$, 
and provides an approximate proof whose soundness bound guarantees
$\|\bdv\|_\infty < \sqrt{q/(2\ell+8)}$.
The former can be done using an inner-product proof system, and the latter can
be done using approximate proof.
One subtlety is that an inner-product proof system binds two a priori
independent committed vectors, one on each generator vector; to prove
$\|\bdv\|_2^2 = \langle \bdv, \bdv \rangle$ the prover additionally shows
that the two committed sides are equal, via the same random-challenge
consistency check used in Equation~\ref{eq:extra-ip} below.

A second subtlety is zero knowledge: a plain inner-product argument is not
zero-knowledge, so the quadratic relations are proven on masked vectors. To
prove $\langle \bdv, \bdv \rangle = B^2$ for $\bdv$ committed with hiding
commitments, the prover first commits to independent uniformly random masks
$\bdd_a, \bdd_b$, one on each generator vector, and to the cross terms $t_1 =
\langle \bdv, \bdd_b\rangle + \langle \bdd_a, \bdv\rangle$ and $t_2 = \langle
\bdd_a, \bdd_b \rangle$; upon a challenge $\chi$, derived by Fiat-Shamir after
these commitments are fixed, both parties run the inner-product argument on
$\bdw_a = \bdv + \chi \bdd_a$ and $\bdw_b = \bdv + \chi \bdd_b$ with public
value $\hat{t} = B^2 + \chi t_1 + \chi^2 t_2$. The vectors $\bdw_a, \bdw_b$ are
independent and uniformly distributed, so the (non-zero-knowledge)
inner-product transcript is simulatable; $\hat{t}$ and the accumulated
commitment blindings are opened homomorphically against the commitments to
$t_1, t_2$, and everything opened is uniform or a function of $\bdw_a,
\bdw_b$. Soundness follows from Schwartz--Zippel on the degree-2 identity
$\langle \bdw_a, \bdw_b \rangle = B^2 + \chi t_1 + \chi^2 t_2$ at the point
$\chi$, whose constant coefficient is $\langle \bdv, \bdv\rangle$ once the two
committed sides are shown equal. The same masking applies verbatim to the
symmetric bilinear relation in Equation~\ref{eq:Linf-ip}.

Proving $L_\infty$ is similar but with a difference that we need to find four
squares for each entry of the vector $\vec{b}\circ \vec{b} - \vec{x}\circ
\vec{x}$ where $\vec{b}:=B\cdot \vec{1}$. The reason we need
$\vec{b}\circ \vec{b} - \vec{x}\circ \vec{x}$ is because we need to
consider both positive and negative entries of $\vec{x}$, and if entries of
$\vec{x}$ fall into $[-B,B]$, then entries of $(\vec{b}+\vec{x})\circ
(\vec{b}-\vec{x})$ are all non-negative and bounded by $B^2$. Namely, there
exist $\vec{a}_1 , \ldots, \vec{a}_4$ such that
\[ \vec{x}\circ \vec{x} + \vec{a}_1 \circ \vec{a}_1 + \cdots + \vec{a}_4 \circ \vec{a}_4 =
\vec{b}\circ \vec{b}. \]
Let $\vec{z}=\vec{x} | \vec{a}_1 | \cdots | \vec{a}_4$; as for $\bdv$,
$\commit(\vec{z})$ is assembled homomorphically from $\commit(\vec{x})$ and a
commitment to $\vec{a}_1 | \cdots | \vec{a}_4$ that the prover publishes before
$r$ is sampled. We then apply Schwartz-Zippel with
$\vec{r}:=(r^0, r^1, \ldots, r^{\ell-1})$ where $r$ is random in $\Z_q$,
\[\langle \vec{r} , \vec{x}\circ \vec{x} + \vec{a}_1 \circ \vec{a}_1+ \cdots + \vec{a}_4 \circ
\vec{a}_4\rangle = \langle \vec{r}, \vec{b}\circ \vec{b} \rangle. \]
We rewrite the above equation as 
\begin{equation}\label{eq:decompose}
   \langle \vec{r} \circ \vec{x}, \vec{x} \rangle
+ \langle \vec{r} \circ \vec{a}_1, \vec{a}_1\rangle 
+ \langle \vec{r} \circ \vec{a}_2, \vec{a}_2\rangle 
+ \langle \vec{r} \circ \vec{a}_3, \vec{a}_3\rangle 
+ \langle \vec{r} \circ \vec{a}_4, \vec{a}_4\rangle
= \langle \vec{r}, \vec{b}\circ \vec{b} \rangle,
\end{equation}
which is exactly an inner product relation: 
\begin{equation}\label{eq:Linf-ip}
  \langle \vec{r} \circ \vec{x} | \vec{r}\circ \vec{a}_1 | \cdots | \vec{r}\circ \vec{a}_4, 
  \vec{x} | \vec{a}_1 | \cdots | \vec{a}_4 \rangle 
= \langle \vec{r}, \vec{b}\circ \vec{b} \rangle. 
\end{equation}
Note that the RHS of Equation~\ref{eq:Linf-ip} is a public value.
The prover then commits to $\vec{t} := \vec{r}'\circ \vec{z}$ where
$\vec{r}'=\vec{r}|\vec{r}|\vec{r}|\vec{r}|\vec{r}$, and proving the relation 
in Equation~\ref{eq:Linf-ip} goes in two steps: first, invoke any inner-product
proof system for Equation~\ref{eq:Linf-ip}, whose commitment is assembled
homomorphically from $\commit(\vec{t})$ and $\commit(\vec{z})$ (the latter
assembled from commitments published before $r$ is sampled), so that the relation binds to the
pre-committed $\vec{z}$; and second, prove that the messages underlying the two
commitments indeed have a linear relation w.r.t. $\vec{r}'$.
For the latter, the prover proves
$\vec{t} = \vec{r}'\circ \vec{z}$ as follows. 
Let $\alpha$ be a random challenge in $\Z_q$, sampled after
$\commit(\vec{t})$ is fixed, and $\bm{\alpha}=(1, \alpha,
\ldots, \alpha^{5\ell-1}) \in \Z_q^{5\ell}$. Then
\[\langle \bm{\alpha}, \vec{t} \rangle 
= \langle \bm{\alpha} , \vec{r}'\circ \vec{z}\rangle
= \langle \bm{\alpha} \circ \vec{r}', \vec{z} \rangle,
\]
and we rewrite as 
\begin{equation}\label{eq:extra-ip}
  \langle \bm{\alpha} | -\bm{\alpha} \circ \vec{r}',
  \vec{t} | \vec{z} \rangle = 0.
\end{equation}
Note that $\bm{\alpha} | -\bm{\alpha} \circ \vec{r}'$ is public. Now we can
invoke a linear proof for Equation~\ref{eq:extra-ip}, with the public vector
being $\bm{\alpha} | -\bm{\alpha} \circ \vec{r}'$ and the secret vector being
$\vec{t} | \vec{z}$; its commitment is obtained from the individual commitments
to $\vec{t}$ and $\vec{z}$. Finally, an approximate proof on $\vec{z}$ whose
soundness bound satisfies $B' < \sqrt{q/10}$ (together with $B^2 < q/2$)
ensures that each entry equation in Equation~\ref{eq:decompose} holds over the
integers, preventing coordinate-wise wraparound.

To summarize, proving $L_2$ norm requires a length-$(\ell+4)$ quadratic proof,
a length-$2(\ell+4)$ linear proof that its two committed sides are equal, and a
length-$(\ell+4)$ approximate proof which can be instantiated using a
length-$(\ell+132)$ linear proof. Proving $L_\infty$ norm requires a
length-$(5\ell)$ quadratic proof, a length-$(10\ell)$ linear proof for the
consistency check in Equation~\ref{eq:extra-ip}, and an approximate proof on a
length-$(5\ell)$ vector. We can reduce the $L_\infty$ proof to three squares
without leaking anything: $4(B^2-x^2)+1$ is odd and $\equiv 1 \pmod 4$, so it is
never of the form $4^a(8b+7)$, and by Legendre's three-square theorem there are
$a_1, a_2, a_3$ with $4x^2 + a_1^2 + a_2^2 + a_3^2 = 4B^2+1$ whenever $|x|\le
B$; conversely, this equation over the integers implies $|x|\le B$. The proof is
as above with $\vec{z} = \vec{x} | \vec{a}_1 | \vec{a}_2 | \vec{a}_3$,
$\vec{r}' = 4\vec{r} | \vec{r} | \vec{r} | \vec{r}$ and right-hand side
$(4B^2+1)\langle \vec{r}, \vec{1}\rangle$; as $|a_j| \le 2B+1$, the approximate
proof on $\vec{z}$ needs $\gamma(2B+2) < \sqrt{q/14}$, so that $4x^2 +
a_1^2+a_2^2+a_3^2 < q/2$. The quadratic proof, consistency-check linear proof,
and approximate proof for $L_\infty$ norm then have length $4\ell$, $8\ell$,
and $4\ell$, respectively. (The $L_2$ proof keeps four squares, which add only
four entries.)

\paragraph{Bounding the LWE error.} For decryption, the LWE error $\bde_i$ needs
a bound $B_\bde$ far above what the $L_\infty$ proof above supports (its
approximate proof needs $\gamma(2B+2)<\sqrt{q/14}$). We therefore decompose the
error into two digits as in~\cite[\S3.5--3.6]{GHL22}: $\bde_i = 2^{h}\bde_i^{\mathrm{hi}} +
\bde_i^{\mathrm{lo}}$, where $\bde_i^{\mathrm{hi}}$ is $\bde_i/2^{h}$ rounded
and $\|\bde_i^{\mathrm{lo}}\|_\infty \le 2^{h-1}$. In place of
$\commit(\bde_i)$, the client commits to $\bde_i^{\mathrm{hi}}$ and proves
$\|\bde_i^{\mathrm{hi}}\|_\infty \le B_{\mathrm{hi}}$ with the $L_\infty$ proof
above. As in~\cite[\S3.6]{GHL22}, $\bde_i^{\mathrm{lo}} = \bdy_i -
\bdA\lwesecret_i - 2^{h}\bde_i^{\mathrm{hi}} - \Delta\bdx_i$ is not committed,
since it is fixed by $\bdy_i$ and the commitments; the client proves
$\|\bde_i^{\mathrm{lo}}\|_\infty \le \gamma (2^{h-1}+1)$ with the approximate
proof (which is complete for entries below $2^{h-1}+1$),
whose linear proof also proves the encryption relation
(\appref{app:full-enc-proof}). Unlike~\cite{GHL22}, which proves both digits
exactly, the approximate proof suffices for the low digit, since its gap
$\gamma$ is scaled down by $2^{h}$: the proven bound is $B_\bde =
2^{h}B_{\mathrm{hi}} + \gamma (2^{h-1}+1)$.

Combined with the linear proof of encryption and secret sharing, and using this
three-number optimization for the $L_\infty$ proofs, the overall cost for proof
is summarized as follows.

\begin{lemma}[Cost for proof of encryption and norms]\label{thm:zkp-cost} 
    Given a set of parameters $(\lambda, \ell, q,
    C)$ and let $\lwesecret \in \Z_q^\lambda, \bdx\in\Z_q^{\ell} $. Let $\G $ be a group
    of size $q$. Let $\Delta$ be a constant. Let 
    $$
    \begin{aligned}
    & \mathbb{CS}_{\text{enc}} : \{  \zkio: (\commit(\lwesecret), \commit(\bdx), \commit(\bde^{\mathrm{hi}})), \\
        & \qquad \qquad \zkstate: \bdy =  \bdA \cdot \lwesecret + \bde + \Delta \cdot \bdx, \|\bdx\|_2 \le B_\bdx(L_2) ,\\
        & \qquad \qquad \|\bdx\|_\infty \le B_{\bdx} (L_\infty), \| \bde \|_\infty \le B_{\bde}(L_\infty), \\
        & \qquad \qquad \witness: (\lwesecret, \bdx, \bde) \}. 
    \end{aligned}
     $$
     There exist a commit-and-proof protocol $\Pi_{\text{enc}}$
     (\appref{app:full-enc-proof}) with group $\G$ of order $q$ for proving the
     above statement with the following cost, dominated by the inner-product
     proof (either linear proof or quadratic proof) invocations:  

    \begin{itemize}
        \item Quadratic proofs: 2 length-$(4\ell)$, 1 length-$\ell$, 
        \item Linear proofs: 2 length-$(8\ell)$, 1 length-$(2\ell)$, 1 length-$\ell$, 1
        length-$(2\ell + \lambda)$, 1 length-$(\lambda+C)$,
    \end{itemize}
    where we omit the lower order terms and write e.g., $4\ell+128$ as $4\ell$.
\end{lemma}

\begin{figure*}[!tb]
    \centering
 \resizebox{0.95\textwidth}{!}{
 \begin{tcolorbox}[width=\textwidth, enhanced, boxsep=1mm, left= 0mm, right=0.5mm, title={\textbf{Secure aggregation for training iteration $t$ }}]
    \linespread{1.4}
    
    Server and clients agree on public parameters: 
    \begin{itemize}
        \item LWE parameters $(\lambda, \ell, p, q, \bdA\in\Z_q^{\ell
                \times\lambda})$ and $\Delta = \floor{q/p}$. 
        \item Proof parameters: Let $\G$ be the group of order $q$ for the
        commit-and-proof system. Let $\bold{F}, \bold{G}, \bold{H}, \bold{K}$ be
        vectors of generators in $\G$ of length $\lambda, \ell, \ell, C$. The
        norm bounds are $B_{\bdx}(L_{\infty}), B_{\bdx}(L_2), B_\bde(L_{\infty})$. 
                
        \item System model parameters: dropout rate is $\delta$ and malicious
        rate over this iteration of selected clients is
                 $\eta$.
    \end{itemize}
    \textbf{Setup.} The set $\comset$ of \coms is
    determined independently from the aggregation as described in
    Remark~\ref{rmk:election}, with sharing degree $d$. Let the secret key and
    public key for $j\in \comset$ be $(SK_j, PK_j)$ for the labeled encryption
    $\AsymEnc$ (\S\ref{s:agreement}); $j$ registers $PK_j$ with a proof of
    knowledge of $SK_j$ bound to $j$, and \coms without a valid
    registration are removed from $\comset$.

    \smallskip
    \textbf{Round 1 (Server $\rightarrow$ Clients)}
    
    Server notifies a set $S_t$ of $n$ clients (we write $[n]$ for their
    IDs) to start the iteration with ID $t$. It also
    tells the \coms the IDs of the $n$ clients. Each aggregation has its
    own ID $t$, unique among all aggregations that use the \coms' keys (a retry
    after an abort, a server that restarts, or another training process that
    uses the same \coms gets a fresh one), and a client's index $i$ is its ID,
    fixed across aggregations, so that each label $(t,i,j)$ and each proof context belongs to
    one client. 

    \smallskip
    \textbf{Round 1 (Clients $\rightarrow$ Server)}
   
    Client $i \in S_t$ on input $\bdx_i\in\Z_p^\ell$, computes the following:
   
    \begin{enumerate} 
        \item Compute $\bdy_i = \bdA\cdot\lwesecret_i + \bde_i+  \Delta \cdot
        \bdx_i \bmod q$, where
        $\lwesecret_i \xleftarrow{\$} \Z_q^\lambda, \
        \bde_i \leftarrow \chi^\ell$.
        \Comment{For outer aggregation}
        
        \item Compute degree-$d$ packed secret sharing of $\lwesecret_i$ 
        as $\bm{\share}_i = (\share_i^{(1)}, \ldots,
        \share_i^{(C)})$.
        \Comment{For inner aggregation}
    
        \item Compute commitments $\commit_{\bold{F}}(\lwesecret_i),
        \commit_{\bold{G}}(\bde_i^{\mathrm{hi}}), \commit_{\bold{H}}(\bdx_i)$;
          and 
        $\commit_{K_j}(\share_i^{(j)})$ for $j\in \comset$, where $\bold{K}$ is
        parsed as $\{K_j\}_{j\in \comset}$.

        \item Set constraint system $\mathbb{CS}_{\text{shares}}$: 
        \[ \{ \zkio: (\{\commit_{K_j}(\share_i^{(j)})\}_{j\in\comset}, \commit_{\bold{F}}(\lwesecret_i), \bfw, \bm{\gamma} ),
        \quad \zkstate:  \langle \bm{\share}_i | \lwesecret_i, \bfw \rangle
        = 0,
        \quad \witness: (\bm{\share}_i , \lwesecret_i, \text{commitment randomness}) \} ,\]
            and compute $\pi_{\text{shares}} \leftarrow \Pi_{\text{linear}}.\prover(\zkio, \zkstate, \witness)$, where $\bfw\in\Z_q^{C+\lambda}$ (from the dual code for the evaluation points $(1,\ldots,C,0,-1,\ldots,-\lambda+1)$) and the weights $\bm{\gamma}\in(\Z_q^*)^{C+1}$ are derived by Fiat-Shamir from all these commitments, and $\Pi_{\text{linear}}$ runs on $\commit_i^{\bm{\gamma}} := \commit_{\bold{F}}(\lwesecret_i)^{\gamma_0}\prod_{j\in\comset} \commit_{K_j}(\share_i^{(j)})^{\gamma_j}$ with each entry of $\bfw$ divided by its weight (\S\ref{s:robust}).
        
        \item Set constraint system $\mathbb{CS}_{\text{enc}}$:
        $$
        \begin{aligned}
            \{ & \zkio: (\bdy_i,\commit(\lwesecret_i), \commit(\bdx_i), \commit(\bde_i^{\mathrm{hi}})), \\
               & \zkstate: \bdy_i =  \bdA \cdot \lwesecret_i + \bde_i + \Delta \cdot \bdx_i , 
                \|\bdx_i\|_2\le B_{\bdx} (L_2), \|\bde_i\|_\infty \le B_\bde(L_\infty) , \| \bdx_i \|_\infty \le B_{\bdx}(L_\infty),\\
               & \witness: (\lwesecret_i, \bdx_i, \bde_i) \},
        \end{aligned}
        $$
        and compute $\pi_{\text{enc}} \leftarrow \Pi_{\text{enc}}.\prover(\zkio, \zkstate, \witness)$.  
   
    \item Send a tuple to the server:
    $$
    \begin{aligned}
    \{ & \text{``server''} : (\bdy_i, \commit(\lwesecret_i), \commit(\bdx_i), \commit(\bde_i^{\mathrm{hi}}), \{\commit_{K_j}(\share_{i}^{(j)})\}_{j\in\comset}, \pi_{\text{shares}}, \pi_{\text{enc}}); \\  
       & \text{``\com $j \in\comset$''}  : \mathsf{ct}_{i,j} := \AsymEnc (PK_j, (t,i,j), (\share_i^{(j)}, r_i^{(j)}))   \}
    \end{aligned}
    $$
    \end{enumerate}
    
\end{tcolorbox}
}
    \caption{\sys{} protocol description for computing a single sum privately
    (Part I).}
    \label{fig:full}
\end{figure*}

\begin{figure*}[!tb]
    \centering
 \resizebox{0.95\textwidth}{!}{
 \begin{tcolorbox}[width=\textwidth, enhanced, boxsep=1mm, left= 0mm, right=0.5mm, title={\textbf{Secure aggregation for training iteration $t$ contd.}}]
    \linespread{1.4}

    \smallskip
    \textbf{Round 2 (Server $\rightarrow$ \Coms)}

    Let $\calX_1$ be the clients who sent the prescribed messages in Round 1, including a well-formed $\mathsf{ct}_{i,j}$ for every $j\in\comset$.
    
    The server for each client $i \in \calX_1$ computes:
    \begin{enumerate} 
        \item  Derive $\bfw, \bm{\gamma}$ and compute $\commit_i^{\bm{\gamma}}$ as in Round 1.

        \item  Run $ \Pi_{\text{linear}}.\verifier(\zkio, \zkstate,
        \pi_{\text{shares}})$ and 
               $ \Pi_{\text{enc}}.\verifier(\zkio, \zkstate, \pi_{\text{enc}})$.
        \item Remove all clients with invalid proof from $\calX_1$. Call
        this set $\calX_2$.
        \item For clients in $\calX_2$, forward $\mathsf{ct}_{i,j}$ and $\commit_{K_j}(\share_i^{(j)})$ to $j\in
        \comset$. 
    \end{enumerate}

    \smallskip 
     \textbf{Round 2 (\Coms $\rightarrow$ Server)}

     Each \com $j \in \comset$:
     \begin{enumerate}
        \item If it received messages for fewer than
        $(1-\delta-\eta)n$ distinct clients among the $n$ notified in Round 1, abort. Otherwise continue.
         \item For every forwarded client $i$: if $\mathsf{ct}_{i,j}$ is not a valid ciphertext under the label $(t,i,j)$, create a verifiable complaint that $\mathsf{ct}_{i,j}$ is invalid.
         Otherwise, compute $(\share_i^{(j)}, r_i^{(j)})  := \AsymDec(SK_j, \mathsf{ct}_{i,j})$, and check if they are consistent with
         $\commit_{K_j}(\share_i^{(j)})$. If not, create a verifiable complaint that consists of a proof $\pi_{\text{dec}}$ of the decryption $(\share_i^{(j)}, r_i^{(j)})$ of $\mathsf{ct}_{i,j}$ under $SK_j$ (for TDH2, the key $u^{x_j}$ of $\mathsf{ct}_{i,j}$ with a Chaum-Pedersen proof, from which the server computes $(\share_i^{(j)}, r_i^{(j)})$).
         \item It formed a set $\mathcal{V}_j$ that consists of all the clients whose shares are valid.
     \end{enumerate}

    \smallskip
    \textbf{Round 3 (Server $\rightarrow$ \Coms)}

    The server verifies every complaint by checking that $\mathsf{ct}_{i,j}$
    is not a valid ciphertext under the label $(t,i,j)$, or that it is valid,
    $\pi_{\text{dec}}$ proves its decryption $(\share_i^{(j)}, r_i^{(j)})$, and
    they are inconsistent with $\commit_{K_j}(\share_i^{(j)})$. It tells all the
    \coms the set $\mathcal{B}$ of clients with a valid complaint, and sets
    $\mathcal{S}_3:=\calX_2\setminus \mathcal{B}$.
     
    \smallskip
    \textbf{Round 3 (\Coms $\rightarrow$ Server)}

    Each \com $j\in \comset$: 
    \begin{enumerate}
        \item Use the common final set $\mathcal{S}_3$. If $\mathcal{S}_3 \not\subseteq \mathcal{V}_j$, abort.
        \item Compute $\share^{(j)} := \sum_{i\in \mathcal{S}_3}
        \share_{i}^{(j)}$ and send it to the server.

    \end{enumerate}
       
    Server reconstructs the shares $\{\share^{(j)}\}_{j\in \comset}$ to
    $\lwesecret$ with error correction, and computes $\bdy :=  \sum_{i \in \mathcal{S}_3} \bdy_i$ and computes
    $\lfloor \bdy - \bdA \cdot \lwesecret  \mod q \rceil_{\Delta}$. 
 
\end{tcolorbox}

}

    \caption{\sys{} protocol description for computing a single sum privately
    (Part II).}
    \label{fig:full2}
\end{figure*}

\section{Security analysis}\label{s:security}

In this section, we discuss how to select proper parameters for our protocol,
and formally state the properties of \sys{}.

\paragraph{Parameters.}
The system \sys{} has a set of parameters listed below. First, $n$ is the number
of clients per aggregation (a training iteration). $(\lambda, \ell, p, q)$ are
LWE parameters (for the outer aggregation), $(C, d, \lambda)$ are secret-sharing
parameters (for the inner aggregation), where $C$ is the number of shares (which
equals to the number of \coms), $d$ is the degree of the secret-sharing
polynomial and $\lambda$ is the number of secrets. Thus the packed-sharing
privacy threshold is $d-\lambda+1$. $B_{\bdx}(L_\infty),
B_{\bdx}(L_2), B_\bde(L_\infty)$ are bounds on norms. The parameters $n$ and the
norm bounds depends on the machine learning setting which is orthogonal to
security analysis. We choose $q$ to be a prime that is both the LWE modulus and
the order of the group used by the commit-and-proof system. For $(\lambda, \ell,
q)$, we choose parameters satisfying the LWE/Hint-LWE assumption used in
Definition~\ref{def:hint-lwe}~\cite{APS15, DDGR20, CCLS20}. The choice of $p$ depends
on applications (see concrete numbers in \S\ref{s:eval}).

The choice of $C$ and $d$ is shown below. Recall that in our protocol
(\S\ref{s:overview}), each client secret-shares a vector of length $\lambda$
using a polynomial of degree $d$. We must have
\allowdisplaybreaks
\begin{align*}\allowdisplaybreaks
    & d - \lambda + 1 > C \cdot \eta_\comset \quad \text{by security of packed secret sharing, }\\
    & d < C(1-\delta_\comset) \quad  \text{necessary condition to reconstruct the secret. } \\
    &  2C\cdot \eta_\comset + d < (1-\delta_\comset) C \quad \text{in order to do error correction.}
\end{align*} 
We combine the equations and get 
\begin{equation}\label{eq:d-range}
    C\eta_\comset + \lambda - 1 < d < C(1-\delta_\comset-2\eta_\comset).
\end{equation}
We assume an integer $d$ satisfying these inequalities is fixed. A sufficient
asymptotic condition is $C> (\lambda-1)/(1-\delta_\comset -
3\eta_\comset)$ with enough integer slack to choose such a $d$.

\medskip

Theorem~\ref{thm:main} formally states the properties of \sys{}. 

\begin{theorem}\label{thm:main}
Let $\Phi$ be the protocol in Figures~\ref{fig:full} and~\ref{fig:full2} with
LWE parameters $(\lambda, \ell, p, q)$ for a prime $q$ and Regev error
$\bde=\bde_1+\bde_2$, $\bde_1,\bde_2\leftarrow\chi'^{\ell}$, and let
$\mathcal{Z}$ be the set of inputs $\bdx$ with $\|\bdx\|_2\le B_{\bdx}(L_2)$
and $\|\bdx\|_\infty\le B_{\bdx}(L_\infty)$. Let $(\delta, \eta, \delta_\comset,
\eta_\comset)$ be the threat model parameters of Section~\ref{s:pb}, and let
$\comset$ be sampled at random from the clients before the aggregation. Assume
that
\begin{enumerate}[label=(\roman*), leftmargin=*, nosep]
  \item the (sparse~\cite{JLS24}) Hint-LWE assumption
  (Definition~\ref{def:hint-lwe}) holds for $(\lambda, \ell, q, \chi')$;
  \item the commit-and-proof protocols, the registration proofs and the
  Chaum-Pedersen proofs are zero-knowledge and (knowledge) sound, the
  commitments are hiding and binding, and Fiat-Shamir is modeled by a random
  oracle;
  \item the encryption $\AsymEnc$ to the \coms (\S\ref{s:agreement}) has
  publicly verifiable ciphertexts and is IND-CCA secure with labels and
  decryption shares~\cite{EC:ShoGen98};
  \item $nB_\bde<\Delta/2$, a sum of $n$ inputs in $\mathcal{Z}$ does not wrap
  modulo $p$, and honest clients' inputs lie in $\mathcal{Z}$ and their proofs
  verify except with negligible probability (\S\ref{s:proof-norms}); and
  \item the degree $d$ satisfies~\eqref{eq:d-range}.
\end{enumerate}
Then, under the communication model of Section~\ref{s:pb}:
\begin{itemize}[leftmargin=*, nosep]
  \item $\Phi$ realizes $\mathcal{F}$ (Figure~\ref{fig:ideal-sum}) against a
  static adversary that controls the server semi-honestly and, maliciously,
  an $\eta_\comset$ fraction of $\comset$ and an $\eta$ fraction of the $n$
  clients in each aggregation, also over polynomially many aggregations with
  distinct IDs under the same \coms' keys;
  \item $\Phi$ has dropout resilience (Definition~\ref{def:dropout-resilience});
  \item $\Phi$ has disruption resistance
  (Definition~\ref{def:disruption-resistance}) for
  $f$ the sum and $\mathcal{Z}$.
\end{itemize}
\end{theorem}

\begin{remark}[Privacy against a malicious server]
    \label{sub:mal}
      In a simulation-based proof, to prove privacy against the server, we need to simulate the server's view throughout the protocol execution. 
      When the server is malicious, a challenge in the simulation is that the set of honest online clients is only determined
      after the corrupted server has seen the ciphertexts (crafted by the simulator). Consequently,
      the simulation could rely on the Programmable Random Oracle Model (PROM) to go through, where certain outputs of cryptographic primitives
      are re-programmed after this set has been determined. This is widely used in 
      several prior
      works~\cite{bell20secure,ma23flamingo,kp24opa,EPRINT:BGLRS24} to prove privacy.\footnote{One
      could conceivably avoid this by relying simply on secret-sharing of
      the entire vector. However, this increases the communication to the \coms, which is
      unfavorable.} Our complaint mechanism also relies on
      a semi-honest server: a malicious server could forward an honest client's
      ciphertext with a wrong share commitment, and the honest \com's
      complaint would reveal that ciphertext's key.

      Specifically, their proof requires the honest clients to compute
      $\mathcal{H}(x)=y$ for a secret $x$. Later, the simulator programs the
      choice of this value, changing it from $y$ to $y'$. However, this
      programming is incompatible with input validation where the clients prove
      the correct computation of $\mathcal{H}(x)$. This impossibility stems from
      the fact that if there exists a zero-knowledge proof system to prove that
      for secret witnesses $x,y$ and some cryptographic primitive $\mathcal{H}$
      (usually $\mathcal{H}$ is a PRG or hash function in prior works), then
      there needs to be a commitment to $y$ which would enable the verifier to
      verify that the computation was indeed successful. However, if one were to
      program the value of $\mathcal{H}(x)=y'$ after the generation of the
      proof, the commitment to $y$ now needs to commit to $y'$, violating the
      binding property of commitment schemes, unless the simulator can
      equivocate the commitment (e.g., it knows a trapdoor of the commitment key). 
      
      While prior works with input validation such as RoFL~\cite{lycklama23rofl}
      and ACORN-detect~\cite{bell22acorn} claim to guarantee privacy against a
      malicious server, their proofs cannot follow the above simulation unless the commitments are equivocable; 
      and indeed no formal proof was shown in these works.  
      It is nevertheless important to note that this impossibility result does
      not rule out the existence of maliciously secure aggregation protocols.
    \end{remark}

\begin{remark}
Unlike some prior works~\cite{ma23flamingo}, regular clients in \sys{} do not
  maintain any state across different aggregation executions. As a result, when
  the adversary changes the set of corrupted regular clients after each
  aggregation (each training iteration) completes, the protocol still remains
  private against the server. \Coms, however, hold long-lived decryption keys:
  a \com corrupted in a later iteration could decrypt the ciphertexts sent to it
  earlier, which the server may have kept. Their corruption must therefore be
  static for the lifetime of their keys, or the keys must be rotated and the old
  ones erased.
  Like all the prior works~\cite{bell22acorn, bell20secure, ma23flamingo},
  \sys{} is not proven to provide privacy if the adversary changes corrupted
  set within an aggregation execution.
\end{remark}
\section{Optimizations}\label{s:impl}
\label{s:crypto-instantiation}

\paragraph{Sparse LWE} 
The bulk of server-side decryption lies in computing the matrix-vector product
$\bdA \cdot \lwesecret$. When $\bdA$ is sparse, that is, each row of $\bdA$ has
only $k$ non-zero entries, this computation can be significantly accelerated. To realize this, we can
replace the standard LWE assumption used in Regev's encryption scheme
(\S\ref{s:prelim}) with Sparse LWE assumption~\cite{JLS24}. Its concrete
security is estimated by the size of the smallest dense minor of $\bdA$
(Conjecture 7.2 and the script of Appendix A in~\cite{JLS24}): sparse LWE is at
least as hard as LWE whose dimension is that size. We use $k=32$ non-zero
entries per row and $\lambda=256$; the script gives an equivalent LWE dimension
of 239, 232, 226, 221 and 216 for $m=\ell=832$, 1856, 3904, 8000 and 16192
samples (in general, a secret less than $1.5\times$ longer than the LWE one
suffices for $k\ge 30$ and up to $2^{21}$ samples~\cite[Table 1]{JLS24}).

\paragraph{Multi-exponentiation}  
Naively computing commitments to a length-$m$ vector requires $m+1$ group
exponentiations and $m$ group multiplications (\S\ref{s:prelim}). We can reduce the number of group
exponentiations to sublinear in $m$ using the Pippenger algorithm, given in
Lemma~\ref{lemma:pippenger}.
In short, the Pippenger algorithm requires only a small number of expensive
group exponentiation (e.g., $\sigma$ is typically no larger than 256) by
increasing the cheap group multiplication by a factor of $\sigma$. Therefore, we
can use Pippenger for Pedersen vector commitment in any of our inner-product
proofs.

\begin{lemma}[Complexity of Pippenger algorithm~\cite{pippenger, bootle,
    pippenger-hackmd}]\label{lemma:pippenger} Let $\G$ be a group of order $q
    \approx 2^\sigma$, and $G_1, \ldots, G_m$ be $m$ generators of $\G$. Given
    $v_1, \ldots, v_m \in\Z_q$, Pippenger algorithm can compute $G_1^{v_1}
    \cdots G_m^{v_m}$ using $\frac{2\sigma m}{\log m}$ group multiplications and
    $\sigma$ group exponentiations. 
\end{lemma}

\paragraph{Optimistic batch verification}
There are two verification steps we can optimize. First, each \com verifies
 that each online client $i$'s share $\share_i$ and randomness $r_i$ match the
 commitment $\commit_{K_j}(\share_i)=K_j^{\share_i}H^{r_i}$.
 Doing this individually for each share requires up to $n$ group
 exponentiations; exponentiation is an expensive operation. We can apply
 the batch verification technique~\cite{BGR98} to reduce the number of
 exponentiations to sublinear in $n$. Say the \com has shares $\share_1,
 \ldots, \share_n$ with randomness $r_1, \ldots, r_n$, and it wants to batch verify if they match with
 $\commit_{K_j}(\share_1), \ldots, \commit_{K_j}(\share_n)$ respectively. It samples
 random 128-bit values $\alpha_1, \ldots, \alpha_n$
 and checks if
 \[
 \begin{aligned}
 &K_j^{\alpha_1\share_1 + \cdots + \alpha_n \share_n} H^{\alpha_1 r_1 + \cdots + \alpha_n r_n}\\
 &\qquad = \commit_{K_j}(\share_1)^{\alpha_1} \cdots \commit_{K_j}(\share_n)^{\alpha_n}.
 \end{aligned}
 \]
 The
 batch verification only needs two length-$n$ inner products on $\Z_q$ and a single
 length-$n$ multi-exponentiation. 

The batching technique can also be applied to proof verification at the server.
For this, we open the black box of the inner-product proof system and utilize a
property of Bulletproof~\cite{bunz18bulletproof} that allows much faster batch
verification rather than verifying each proof individually. To verify a quadratic proof
of length $\ell$, the verifier computation can be abstracted as\footnote{The
linear proof can be abstracted the same way with a length-$\ell$ vector
$\bold{G}$.}  
\begin{equation}\label{eq:bulletproof-verification}
\bold{G}^{\bold{z}} \stackrel{?}{=} \bold{P}^{\bold{r}},
\end{equation} 
where $\bold{G}$ is a vector of group generators of length-$2\ell$ and
$\bold{P}$ is the proof transcript which consists of $2\log\ell$ group elements.
The exponents $\bold{z}, \bold{r}$ only depend on the verifier challenges.
Verifying $n$ quadratic proofs naively would require the server to compute $n$
multi-exponentiation of length $2\ell$ (LHS
of~\ref{eq:bulletproof-verification}), and $n$ multi-exponentiation of length
$2\log\ell$ (RHS of~\ref{eq:bulletproof-verification}).
 
The key idea is that the LHS of~\ref{eq:bulletproof-verification} can be
 computed for a batch of $n$ proofs using a single multi-exponentiation instead
 of $n$ multi-exponentiations; the insight is that all the proofs share the
 same generators $\bold{G}$. To batch verify $n$ proofs with exponents
 $\bold{z}_1, \ldots, \bold{z}_n$, the verifier (server) can sample random
 $\alpha_1, \ldots, \alpha_n $ and computes $\bold{z}' = \alpha_1 \bold{z}_1 +
 \cdots + \alpha_n \bold{z}_n$ and checks if 
\begin{equation}\label{eq:bulletproof-batch}
    \bold{G}^{\bold{z}'} \stackrel{?}{=} \bold{P}_1^{\alpha_1 \bdr_1} \cdots \bold{P}_n^{\alpha_n \bdr_n},
\end{equation}
where $\bold{P}_i$ and $\bdr_i$ are from the RHS of
equation~\ref{eq:bulletproof-verification} for client $i$'s proof. For LHS
of~\ref{eq:bulletproof-batch}, the server computes a single multi-exponentiation
of length $2\ell$. For RHS of~\ref{eq:bulletproof-batch}, the server computes a
single multi-exponentiation of length $2 n\log \ell$. Computing the exponents
$\bold{z}_i$ still takes $O(\ell)$ operations in $\Z_q$ per proof, which dominate the
server's cost for large $n$ (\S\ref{s:eval-bench}).

In both the above cases, if there exist malicious clients such that the batch
 verification returns false, then the \com (or the server) can divide the
 shares (or the proofs) into smaller groups and recursively apply batch
 verification until it identifies the malicious clients; our implementation
 simply verifies them one by one in that case.

\section{Evaluation}\label{s:eval}

In this section, we provide benchmarks to answer the following questions:
\begin{itemize}
    \item What are \sys{}'s concrete costs of the clients and the
          server, for aggregation and proofs, respectively?
    \item What is \sys{}'s overall cost including input validation?
    \item What is the cost of the \coms and how does it compare to the cost 
          of regular clients?
    \item How does \sys{} compare to prior robust secure aggregation
          protocols in terms of costs?
\end{itemize}

\paragraph{Selecting a proper baseline.}
The most relevant work is ACORN-robust~\cite{bell22acorn}: they provide the same
input validation but achieve robustness in a very different way. We provide
details in \appref{app:baselines:acorn-robust} but roughly, ACORN-robust
follows the pairwise masking approach in Bell et al.~\cite{bell20secure}, but
they additionally have a dispute protocol to iteratively find cheating clients
and remove their inputs from the aggregation result. The other prior work with
disruption resistance is Eiffel~\cite{CCS:CGJM22}, but their per-client work is
$O(n^2\ell)$ which is not computationally feasible for mobile clients. Other
works like RoFL~\cite{lycklama23rofl}, ACORN-detect~\cite{bell22acorn} have
strictly weaker property: the server has to abort the aggregation once a
malicious client has been detected. Therefore, we identify ACORN-robust as the
only valid baseline.

\paragraph{Libraries and testbed.} 
We implement our protocol using Rust. We instantiate the proof using
Bulletproof~\cite{bunz18bulletproof, dalek-bulletproof} since it is designed for inner-product relation. 
We use the
dalek library~\cite{dalek-bulletproof} for elliptic curve cryptography.  We run
our experiments on an Apple M4 (4 performance and 4 efficiency cores, 16\,GB
memory). 

\paragraph{Concrete parameter selection.}
We use a proof system based on Ristretto group~\cite{bunz18bulletproof,
dalek-bulletproof} and set the LWE modulus $q$ equal to the group order which is
a 253-bit prime.\footnote{One could use a LWE modulus that is much smaller than
the group order of the proof system, but this requires additional $L_\infty$
proofs and non-trivial changes for the Schwartz-Zippel compressing technique.} 
To ensure correctness of decryption, the accumulated noise from all clients must
remain within the tolerance of the LWE scheme. Concretely, we require $n \cdot
B_\bde < \Delta/2$, where $B_\bde$ is the per-client error bound proven by the client (\S\ref{s:proof-norms}) and $\Delta
\approx q/p$ is the scaling factor for ciphertext modulus $q$ and plaintext
modulus $p$. That is, $2 p n B_\bde < q$. Having fixed $p$ and $n$ based on
applications (e.g., $p$ depends on gradient quantization range), and $q$ to be
the group order of the proof system, we can derive the bound for $B_\bde$; then
knowing $p,q,B_\bde$ allows us to choose the LWE secret length to get
certain security level. In this work, we set $p=2^{16}$, and $n\le 5000$, which
is sufficient for federated learning applications~\cite[Table
2]{kairouz21advances}. Each input entry has absolute value at most 6 (so the sum
of $n$ inputs does not wrap modulo $p$), and the $L_2$ norm of each input is at
most 512. We set $\lambda$ to be 256, $\lwesecret$ uniform in
$\Z_q$, $\bdA$ sparse with $k=32$ non-zero entries per row (\S\ref{s:impl}), and
the error $\bde=\bde_1+\bde_2$ with $\bde_1,\bde_2$ discrete Gaussian of
standard deviation $2^{218}$ (so $\bde$ has standard deviation
$\sigma_\bde=2^{218.5}$), cut at $14\sigma_\bde$. With $h=193$ and
$B_{\mathrm{hi}}=\lceil 14\sigma_\bde/2^{h}\rceil+1$, the proven bound
$B_\bde\approx 2^{222.3}$ satisfies $2pnB_\bde<q$ for $n\le 5000$ and $\ell\le
2^{14}$. By the estimate in~\cite{JLS24} (Conjecture 7.2), this sparse LWE
instance is as hard as LWE of dimension 216, which has 121 bits of classical
core-SVP security ($2^{0.292\beta}$, primal and dual
attacks~\cite{USENIX:ADPS16}) for the Hint-LWE error of standard deviation
$\sigma_\bde/2$ (Definition~\ref{def:hint-lwe}), and 131 bits without the hint;
the security is per key, as $\bdA$ is shared by the clients of an iteration.

\sys{} is designed for federated learning on small-scale models (e.g., mobile or
edge models). Supporting very large models, such as large language models, is
out of scope, since training such models in a federated learning setting is
generally impractical. In the evaluation, we set client input length up
to roughly $2^{14}$, which covers the small-model settings in our benchmarks.

Following prior works~\cite{bell22acorn} we use
power-of-2 input length for benchmarking. Since the inner-product proof system
works efficiently for vectors of length power-of-2, our client's input length is
set slightly smaller than powers of 2 such that the resulting vector input to
the inner product proof system has length exactly or almost power of 2 (nearly
no waste for padding).

\smallskip 

\paragraph{Central theme in evaluation.}
Experiments in the following sections will substantiate a central argument we
make: while \sys{} has a computational cost close to that of ACORN-robust
(\S\ref{s:eval-bench}), the reduction in round complexity plays a crucial role
in lowering the end-to-end runtime. The key to this improvement lies in the
interplay between the allowed dropout rate and server waiting time
(\S\ref{s:eval-round}). 

\subsection{Computation and communication}\label{s:eval-bench}

\paragraph{Regular client costs.} 
We present the breakdown of a regular client's computational cost in
Figure~\ref{fig:client-computation}, varying the input length. The cost for
\coms is discussed separately later. Since our baseline ACORN-robust shares
similar types of computation with \sys{}, we present the costs for two systems
jointly, dividing the process into four steps: input-independent secret sharing,
input masking, commitment generation, and proof generation.\footnote{In
ACORN-robust, each client establishes input-independent pairwise secrets with
$O(\log n)$ neighbors, expands them to $O(\log n)$ masks of input length, and
performs Feldman secret sharing of the secrets. Microbenchmarks for ACORN-robust
assume 40 neighbors per client (from~\cite{bell20secure}).  } In each phase, we
indicate whether ACORN-robust and \sys{} perform the same or different
operations: shared operations are shown in black text, while differences are
highlighted using distinct colors. Since the authors of ACORN-robust did not
implement this protocol, we estimate their client's costs by extracting the
types of operations and counting the number of operations for each type (e.g.,
how many scalar multiplication on elliptic curve), and simulating them with Rust
(the same language used for implementing ours). The client secret-sharing cost
in \sys{} depends on the committee size, and we choose $C=2\lambda$ for the
following reason: recall that $C$ must be larger than $(\lambda-1)/(1 -
\delta_\comset - 3\eta_\comset)$, and since the overall dropout and corruption
rates are small (\S\ref{s:pb}), the sampled committee inherits similarly small
rate; as a result, choosing $C=2\lambda$ already leaves sufficient slack to
satisfy the condition. Concretely, for overall corruption and dropout rates of
at most 5\%, $C=512$ with degree $d=331$ satisfies condition (v) of
Theorem~\ref{thm:main} except with probability $2^{-52}$.

For both systems, the bulk of the computation time is spent on proof generation;
most other phases are only on the order of milliseconds. The $L_2$ and
$L_\infty$ proofs (for input validation) are the same in both systems. \sys{}
has additional work for the proofs its design adds (enc linear, the $L_\infty$
norm of $\bde^{\mathrm{hi}}$ and the scrape test): 0.49 seconds for
$\ell\approx 2^{10}$ and 7.0 seconds for $\ell\approx 2^{14}$, but as
we show later (\S\ref{s:eval-round}), this is a tradeoff that yields significant
gains: while our clients spend more time, the number of rounds for us
is significantly fewer than theirs. Since round complexity is substantially
important for run time in a setting where clients drop out arbitrarily
(evidenced in \S\ref{s:eval-round}), this reduction leads to much overall
performance gains that is far outweigh the effect of the increased
client computation. 

A final complication is that 
ACORN-robust has an additional cheater identification phase to remove
the effects of malicious clients from the aggregation result. Since this 
phase involves only sending small stored messages to the server
(Algorithm 4 in~\cite{bell22acorn}) and there is no cryptographic computation at clients,  
we do not depict them in Figure~\ref{fig:client-computation}.

\begin{figure}[!tb]
    \centering
    \resizebox{\linewidth}{!}{
    \begin{tabular}{@{}lrrrrr@{}}
    \toprule
    Input length $\ell$ approx.  & $2^{10}$             & $2^{11}$             & $2^{12}$             & $2^{13}$             & $2^{14}$             \\ \midrule
    {\color{brown} Packed sharing in {\small Arma.} (ms)} & {\color{brown} 21.58} & {\color{brown} 21.58} & {\color{brown} 21.58} & {\color{brown} 21.58} & {\color{brown} 21.60} \\
    {\color{brown} Compute $\bdA\lwesecret$ in {\small Arma.} (ms)} & {\color{brown} 5.25} & {\color{brown} 11.70} & {\color{brown} 24.65} & {\color{brown} 50.61} & {\color{brown} 103.02} \\
    {\color{gray} Feldman in {\small{ACORN}} (ms)} & {\color{gray} 76.49} & {\color{gray} 76.49} & {\color{gray} 76.49} & {\color{gray} 76.49} & {\color{gray} 76.49} \\ \midrule
    {\color{brown} Masking in {\small Arma.} (ms)} & {\color{brown} 2.96} & {\color{brown} 6.60} & {\color{brown} 13.69} & {\color{brown} 28.06} & {\color{brown} 57.11} \\
    {\color{gray} Masking in {\small{ACORN}} (ms)} & {\color{gray} 0.84} & {\color{gray} 1.89} & {\color{gray} 3.96} & {\color{gray} 8.11} & {\color{gray} 16.39} \\ \midrule
    Commitment (ms) & 22.97 & 24.19 & 26.67 & 31.73 & 41.99 \\
    - commit to $\lwesecret$, $\bdx$, $\bde^{\mathrm{hi}}$ & 3.09 & 4.31 & 6.79 & 11.85 & 22.11 \\
    - {\color{brown} commit to shares in {\small Arma.}} & {\color{brown} 19.88} & {\color{brown} 19.88} & {\color{brown} 19.88} & {\color{brown} 19.88} & {\color{brown} 19.88} \\
    - {\color{gray} commit to masks in {\small{ACORN}}} & {\color{gray} 84.02} & {\color{gray} 157.54} & {\color{gray} 305.72} & {\color{gray} 603.74} & {\color{gray} 1203.08} \\ \midrule
    Proofs (sec)          & \multicolumn{1}{l}{} & \multicolumn{1}{l}{} & \multicolumn{1}{l}{} & \multicolumn{1}{l}{} & \multicolumn{1}{l}{} \\
    - $L_\infty$ norm of $\bdx$ (\S\ref{s:proof-norms}) & 0.38 & 0.74 & 1.45 & 2.91 & 5.83 \\
    - $L_2$ norm of $\bdx$ (\S\ref{s:proof-norms}) & 0.14 & 0.27 & 0.53 & 1.04 & 2.06 \\
    - {\color{brown} $L_\infty$ norm of $\bde^{\mathrm{hi}}$} (\S\ref{s:proof-norms}) & {\color{brown} 0.39} & {\color{brown} 0.79} & {\color{brown} 1.56} & {\color{brown} 3.13} & {\color{brown} 6.23} \\
    - {\color{brown} enc linear} (\S\ref{s:robust}) & {\color{brown} 0.05} & {\color{brown} 0.09} & {\color{brown} 0.17} & {\color{brown} 0.33} & {\color{brown} 0.68} \\
    - {\color{brown} scrape test$^{*}$} (\S\ref{s:robust}) & {\color{brown} 0.05} & {\color{brown} 0.05} & {\color{brown} 0.05} & {\color{brown} 0.05} & {\color{brown} 0.05} \\ \bottomrule
    \end{tabular}
    } \caption{Computation cost per client in \sys{} and ACORN-robust~\cite{bell22acorn} varying input vector
    lengths, benchmarked using a single thread of execution on an Apple M4. Masking in \sys{} includes
    sampling the LWE error; enc linear includes the bound on $\bde^{\mathrm{lo}}$ (\S\ref{s:proof-norms});
    $^{*}$the scrape test includes committing to the shares, also listed under Commitment.}
    \label{fig:client-computation}
\end{figure}

\paragraph{\Com costs.}
Per-\com cost is independent of the input length $\ell$, so we vary the number
of clients $n$ and show the corresponding costs.
For checking and decrypting the ciphertexts and batch verification of the
shares (\S\ref{s:impl}), per-\com computation time is linear to the number of
clients: for 500, 1K, 1.5K and 2K clients, the time is 0.04, 0.08, 0.11 and
0.15 seconds respectively. A complaint about an invalid ciphertext is
checked from the ciphertext alone, and one about a valid ciphertext carries its
decryption key and a Chaum-Pedersen proof (\S\ref{s:agreement}); the server
verifies them in 51 and 138\,$\mu$s respectively, and honest \coms
only complain about corrupted clients. On the client side, encrypting the $C$
shares takes 29\,ms. 
Importantly, these numbers show that per-\com computation is cheaper than that
of the regular client when the regular client's norm-proof cost is included;
excluding norm proofs, the two costs can be comparable for small input lengths
or large $n$.

\begin{figure}[!tb]
\centering
    
    \includegraphics[width=\linewidth]{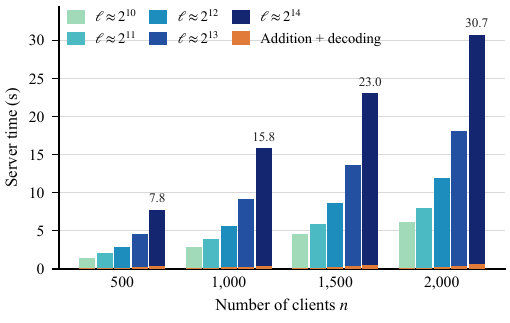}
  \caption{Server computation in \sys{} for different number of clients
  (indicated via x-axis) and different lengths of inputs (bar colors); each bar
  stacks proof verification on addition and decoding, with the total for
  $\ell\approx 2^{14}$ on top. Benchmarked using 8 threads on an Apple M4
  (4 performance and 4 efficiency cores), verifying $n$ copies of one
  client's submission; the Reed-Solomon decoding is measured on one thread.} 
  \label{fig:server-costs}
\end{figure}

\paragraph{Server costs.} 
\sys{}'s server computation depends on both input length $\ell$ and the number
of clients $n$. Figure~\ref{fig:server-costs} breaks down the computation into
aggregation (summing masked vectors and decoding using $\bds$, including the
error correction of the \coms' shares) and proof verification. Proof
verification dominates; with batching (\S\ref{s:impl}), each additional client
adds about 3\,ms for $\ell\approx 2^{10}$ and 15\,ms for $\ell\approx 2^{14}$. 

For ACORN-robust, the server performs the same proof verification and vector
additions as in \sys{}, and additionally verifying Feldman commitments for every
pairwise seed. Without optimization, Feldman verification could become a
bottleneck: for 1K clients with 40 neighbors each, the server computes $40^2
\cdot 1000$ group exponentiations, taking 40 seconds. 

\paragraph{Communication.}
A regular client sends $\bdy_i$ ($32\ell$ bytes), its commitments and proofs
(41--43\,KB, nearly independent of $\ell$), and $C$ ciphertexts for the
\coms (98\,KB: 192 bytes each and $C=512$): 166\,KB in total for $\ell\approx
2^{10}$ and 660\,KB for $\ell\approx 2^{14}$. Each \com receives one
ciphertext and one share commitment (224 bytes) per client in $\calX_2$, and
sends one element of $\Z_q$ in Round 3.

\subsection{Simulating communication rounds}\label{s:eval-round}

\paragraph{Why rounds matter}
When executing an interactive protocol in the server-client setting, fixing the
server's waiting time per round and dropping any clients who are late is a
common strategy. This exacerbates the impact of round complexity: say a protocol
is theoretically designed to tolerate 15\% dropouts over all rounds of an
aggregation, if this protocol has many rounds, e.g., 20 rounds, it may fail in
practice because even 1\% dropout per round can accumulate to 20\% dropouts in
total in which case the protocol just fails (the server does not get any
aggregation result). This effect is depicted in the first two graphs in
Figure~\ref{fig:message-distribution}: we set the waiting time per round to be 12
seconds and simulate the message arrivals of 1K clients in ABIDES~\cite{abides},
and we plot the arrival times of all messages and count the percentage of late
messages (dropouts). In this case, a 3-round protocol that can tolerate up to
10\% overall dropouts can successfully output the aggregation result, but a
9-round protocol that can tolerate up to 10\% overall dropouts will fail. To
ensure the latter protocol succeeds, one needs to increase the server waiting
time per round, thereby reducing the dropouts per round and making the overall
dropouts within the acceptable threshold. This is shown in the third graph in
Figure~\ref{fig:message-distribution}. As a result, more rounds in fact
amplifies the overall run time increase by increasing the per-round waiting
time.

\begin{figure}[!tb]
\centering
\begin{tabular}{c|c|c|c}
\hline
\textbf{$n$ (clients)} & $\eta=0.01$ & $\eta=0.05$ & $\eta=0.10$ \\
\hline
500  & 9 & 10 & 12 \\
1000 & 9 & 11 & 13 \\
1500 & 9 & 11 & 13 \\
2000 & 9 & 11 & 13 \\
\hline
\end{tabular}
\caption{Number of rounds for ACORN-robust such that, with probability at least
$0.9$, the protocol terminates within this bound. Since the round complexity is
a random variable, the table reports the $0.9$-quantile of its distribution.
\sys{} has 3 rounds for all the above listed $n$ and $\eta$. }
\label{fig:acorn-rounds-eta}
\end{figure}

\begin{figure}[!tb]
    \includegraphics[width=0.36\textwidth]{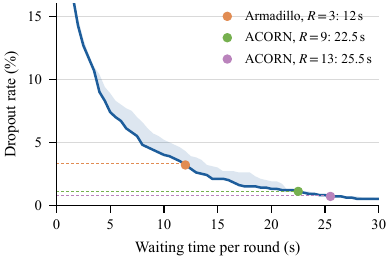}
    \caption{The relation between server waiting time and the dropout rate
    $\hat{\delta}$ of one round trip between the server and 1K clients in the
    ABIDES simulator~\cite{abides}, with the pairwise latency model of
    Flamingo~\cite{ma23flamingo} (median over 5 runs; the band reaches the
    maximum). Points mark the waiting time each protocol needs for
    $\hat{\delta}=\delta/R$ with $\delta=0.1$.}
    \label{fig:waiting-delta}
\end{figure}

\paragraph{Simulation methodology.}
For a fixed dropout rate $\delta$ that a $R$-round protocol can tolerate, we
assume per round the server can handle $\hat{\delta} = \delta/R$ dropout rate.
We derive the relation between waiting time and dropout rate
(Figure~\ref{fig:waiting-delta}) by simulating one round trip between the
server and 1K clients in ABIDES~\cite{abides}. Next, for \sys{} and our
baseline ACORN-robust, given $\delta$ and $R$, we calculate the average
per-round dropout rate $\hat{\delta}$ and determine server waiting time $T$
using Figure~\ref{fig:waiting-delta}. The total round trip time is estimated as
$T \cdot R$, and the total runtime includes this plus the computation time. We
check these estimates by simulating complete executions of both protocols in
ABIDES with these waiting times, the measured computation
(Figures~\ref{fig:client-computation} and~\ref{fig:server-costs}) and message
sizes (\S\ref{s:eval-bench}); in \sys{}, the server verifies the proofs while it
waits for the Round 1 messages.

\begin{figure}
    \includegraphics[width=\columnwidth]{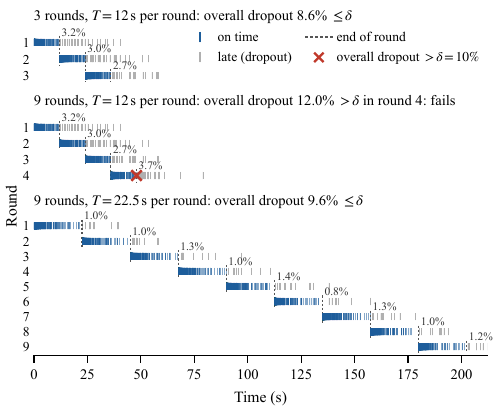}
    \caption{Illustration of the relation between waiting time and dropout
    percentage. With a fixed waiting time per round (12 or 22.5 seconds), blue marks
    denote messages that arrive on time and grey marks those that arrive late;
    the server processes only the blue messages, treating the grey ones as
    dropouts. A protocol fails once its overall dropouts exceed 10\%. Message
    arrivals are simulated in ABIDES~\cite{abides} with 1K clients.} 
    \label{fig:message-distribution}
\end{figure}

\paragraph{Rounds and run time comparison.} 
Table~\ref{fig:acorn-rounds-eta} reports the number of rounds within which the
ACORN-robust protocol terminates with probability at least $0.9$, for different
values of $n$ and $\eta$ (cf.~\cite[Theorem~4.1]{bell22acorn}). Even in the
most favorable setting ($n=500$, $\eta=0.01$), ACORN-robust requires 9 rounds,
which is already three times that of \sys{}, which completes in 3 rounds. For
larger values of $\eta$, the required number of rounds grows further (e.g.,
12--13 rounds when $\eta=0.10$), while \sys{} remains fixed at 3 rounds.
A larger $\eta$ only requires a larger committee in \sys{}: with
$\delta=0.1$, condition (v) of Theorem~\ref{thm:main} fails with probability
below $2^{-52}$ with $C=512$, $560$ and $832$ for $\eta=0.01$, $0.05$ and
$0.10$, which does not change its waiting time. (The simulation uses the
costs measured with $C=512$ for all $\eta$; with $C=832$, a client encrypts and
sends $320$ more shares and each \com's work per client is unchanged.)
Moreover, ACORN-robust involves all clients in every round, whereas \sys{}
contacts all clients only in the first round and then interacts only with the
committees in the remaining two.

\begin{figure}[!tb]
    \includegraphics[width=\linewidth]{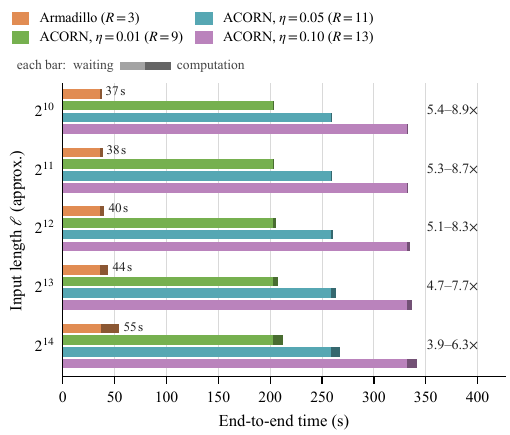}
    \caption{End-to-end time of \sys{} and ACORN-robust in simulated executions
    in ABIDES (\S\ref{s:eval-round}) with 1K clients and $\delta=0.1$, split into
    waiting and computation. ACORN-robust needs $R=9, 11, 13$ rounds for
    $\eta=0.01, 0.05, 0.10$ (Figure~\ref{fig:acorn-rounds-eta}), and each round
    waits until the dropout rate is $\delta/R$ (Figure~\ref{fig:waiting-delta}):
    22.5--25.5 seconds per round, against 12 seconds for each of \sys{}'s 3
    rounds. The numbers next to each group are \sys{}'s speedup. For
    ACORN-robust, only client computation is included, so its time is a lower
    bound.}
    \label{fig:round-time}
\end{figure}

Figure~\ref{fig:round-time} shows that even when excluding client- and
server-side computation, the time spent on waiting and round synchronization
already differs by 5.6--9.2$\times$ between \sys{} and ACORN-robust.
For 1K clients, ACORN-robust incurs 203--332 seconds of waiting compared to only
36--36.5 seconds in \sys{}. Client
computation (Figure~\ref{fig:client-computation}) ranges from about 1.1 seconds
for input length $2^{10}$ to 15 seconds for input length $2^{14}$, and the server
verifies the proofs while it waits for Round 1. As a result, waiting time is the dominant
factor in end-to-end runtime: the simulated executions take 37--55 seconds in
\sys{} and at least 203--341 seconds in ACORN-robust (without its server
computation), a 3.9--8.9$\times$ improvement when total
time including computation is considered. 

Another observation is that since $L_\infty$ proof dominates computation
(Fig.~\ref{fig:client-computation}), if there exists faster proof of norms
(especially proof of $L_\infty$) in the future, it would amplify \sys{}'s
advantage as the portion of round trip time becomes more significant.

\section{Related work and discussion}

In this section we discuss other recent works to compute private sums for
federated learning. 

Bonawitz et al.~\cite{bonawitz17practical} introduced the first secure
 aggregation protocol for federated learning applications. Subsequent
 works~\cite{bell20secure, so22lightsecagg, stevens22efficient, ma23flamingo,
 guo22microfedml, li23lerna,kp24opa,EPRINT:BGLRS24, superfl24zhao,
 bangalore23flag} refine the design to better fit practical federated learning
 deployments. In particular, they 
 \begin{myitemize2}
    \item reduce concrete computation costs at the clients and the
    server~\cite{bell20secure, so22lightsecagg, stevens22efficient,
    ma23flamingo, li23lerna, guo22microfedml, elsa23rathee};
    \item introduce mechanisms to prevent malicious clients from submitting
    malformed model weights~\cite{bell22acorn, lycklama23rofl, elsa23rathee, bangalore23flag}; 
    \item support asynchronous client
    participation~\cite{kp24opa,
    EPRINT:BGLRS24};
    \item robust against arbitrary malicious clients'
    behaviors~\cite{bell22acorn}.
 \end{myitemize2}

In what follows, we focus on their robustness properties---how well these
protocols tolerate malicious behaviors---and organize the discussion
into two categories: protocols in the single-server setting and those in the
multi-server setting.

\paragraph{Single-server setting.}
A few works~\cite{CCS:CGJM22, lycklama23rofl, bell22acorn, EPRINT:BGLRS24}
achieve weaker robustness notion than the one in this work,
specifically, they ensure the server either obtains a valid sum or the protocol
aborts. Notable works in this line include Eiffel~\cite{CCS:CGJM22}, which uses
SNIP~\cite{corrigan-gibbs17prio}, a proof system that allows arbitrary
predicates to be verified over secret-shared inputs by multiple parties. In
Eiffel, each client secret-shares its input vector with all other clients, who
act as the verifiers in SNIP. This incurs high communication overhead: the
per-client communication cost scales as $O(n^2 \ell)$, where $n$ is the number
of clients and $\ell$ is the input length. A later work
RoFL~\cite{lycklama23rofl} builds on the protocol by Bonawitz et al., and requires
clients to provide range proof on their inputs. This approach incurs
significantly lower communication cost compared to Eiffel. Both works do not
guarantee that the honest server always gets the correct sum when some clients
act maliciously.

The most relevant work to ours is the ACORN family~\cite{bell22acorn},
specifically ACORN-detect and ACORN-robust. ACORN-detect reduces proof size
significantly compared to RoFL. ACORN-robust extends this to a more robust
protocol with a probabilistic cheater identification mechanism that requires
$O(\log n)$ rounds and $\polylog(n)$ work per client.

A recent work by Alon et al.~\cite{GMPC22} shows that any function can be
securely computed with $\polylog(n)$ work per client, $\poly(n)$ work at the
server, and $\polylog(n)$ rounds, with guaranteed output delivery, assuming
fully homomorphic encryption exists. However, it remains unclear whether this
general protocol offers advantages over specially designed protocols for
computing sums.

\paragraph{Multiple non-colluding servers.} 
Some works distribute trust across multiple servers, such as the two-server
solutions Elsa~\cite{elsa23rathee} and SuperFL~\cite{superfl24zhao}, or the
generic multi-server solution Flag~\cite{bangalore23flag}. These approaches
involve clients secret-sharing their inputs with multiple servers, which then
communicate to validate inputs. Some other protocols like
Mario~\cite{EPRINT:NguTriLep24} use threshold additively homomorphic encryption
to encrypt inputs for a set of servers. While some of these solutions are faster
in run time compared to single-server protocols, they face criticism for relying
on the non-collusion assumption among servers, which is difficult to ensure when
the same organization owns the servers. 

\ifcamera\else
Our model (and~\cite{bell22acorn, ma23flamingo}) differs fundamentally from the
multi-server approach. In multi-server setups, powerful servers can handle heavy
computation, such as the $n\ell$ work required by secret-sharing
solutions~\cite{elsa23rathee, bangalore23flag}, where $n$ is the number of
clients and $\ell$ is the input length. In single-server models, computational
and communication overhead must be minimized for the resource-constrained
clients. In other words, we must push the heavy computation to the server---we
cannot have a protocol with $n\ell$ work at any client. 

\sys{} (and~\cite{bell22acorn, ma23flamingo}) works by each client sending its masked
input of length $\ell$ to the server; then, in the rest of the steps, the clients
do computation {\em independent} of $\ell$ to help the server unmask the sum.
For ACORN family protocols~\cite{bell20secure, bell22acorn}, each client communicates with
$\polylog(n)$ other clients where each of them doing $\polylog (n)$ work.
For \sys{} (and~\cite{ma23flamingo}), each client communicates
to a set of $\polylog(n)$ clients, where the latter does linear work in $n$. In
practice, $n$ is smaller than the input size $\ell$ so that the input size will
dominate the client cost. 
\fi 

\subsection{Other security considerations}
\paragraph{Set selection among iterations}
So et
al.~\cite{so23securing} demonstrate that the server can infer client data by
observing sums over many training iterations if the subset of clients in each iteration are not carefully chosen. 
They propose a partitioning
strategy to mitigate this risk.
Pasquini et al.\cite{pasquini22eluding} show an attack where a malicious server
can circumvent secure aggregation by sending inconsistent models to clients. To
mitigate this, some works~\cite{pasquini22eluding, ma23flamingo} suggest binding
the model hash to pairwise masks, which cancel out if all clients share the same
hash. Since our secure aggregation protocol does not rely on pairwise masking,
we can use a different approach: each client hashes the received model and sends
the hash to the \coms. The \coms then perform a majority vote on the hashes
and exclude shares from clients whose hashes do not match the majority.

\paragraph{Defense mechanisms}
Norm-based defenses remain a widely adopted and practically effective mechanism.
While more sophisticated federated learning attacks can evade norm-based
safeguards~\cite{fang2020local,xie2024poisonedfl, bernstein2019signsgd,
damaskinos2018asynchronous}, norm clipping or norm bounding nevertheless
provides a highly pragmatic first line of defense against many straightforward
attacks. For example, the experiments in RoFL~\cite{lycklama23rofl} show that
enforcing $L_2$ and $L_\infty$ constraints on client updates can effectively
mitigate many classes of severe attacks. Another empirical
analysis~\cite{sun2019can} in the backdoor attack context shows that setting a
norm bound at the 90th percentile of benign updates can almost mitigate boosted
attacks. These findings underscore the real-world utility of norm bounds: they
block a large swath of naïve or easily executed attacks with minimal cost or
deployment complexity.

\subsection*{Acknowledgments}
We thank James Bell and Adrià Gascón for answering our questions about
ACORN-robust protocol. We thank Jess Woods for helpful discussion on proof
systems and Cathie Yun for answering questions about delak cryptography library.
We thank Guohao Li for pointing out typos in the equations in
Section~\ref{s:proof-norms}.
 We thank
Sebastian Angel, Raluca Ada Popa and Tal Rabin for their support during this
work. Yiping Ma was supported by a
Microsoft Research PhD Fellowship. 

This paper was prepared in part for information purposes by Artificial Intelligence
Research Group and the AlgoCRYPT CoE of JPMorgan Chase \& Co and its
affiliates (“JP Morgan”) and is not a product of the Research Department of JP
Morgan. JP Morgan makes no representation and warranty whatsoever and disclaims
all liability, for the completeness, accuracy, or reliability of the information
contained herein. This document is not intended as investment research or
investment advice, or a recommendation, offer, or solicitation for the purchase
or sale of any security, financial instrument, financial product, or service, or
to be used in any way for evaluating the merits of participating in any
transaction, and shall not constitute a solicitation under any jurisdiction or
to any person, if such solicitation under such jurisdiction or to such person
would be unlawful.

\balance
\bibliographystyle{ACM-Reference-Format}
\bibliography{conferences, paper, abbrev2, crypto}

\ifcamera\else
\appendix

\section{Deferred material for proof of norms}
\label{app:proof-norms}
 
\subsection{Approximate proof}\label{app:approx}
We describe the protocol in Gentry, Halevi, and Lyubashevsky~\cite{GHL22} below. 
Let the security parameter be $\sigma$. 
The prover has a vector $\bda $ of length $m$ where $\|\bda\|_{\infty} < B $. 
Let $B'$ be the bound that the prover can prove with the following protocol.
For security, the gap $\gamma := B'/B$ should be larger than $27.56 \sigma \sqrt{m}$.
Set the rejection thresholds as
\[ \tau_{\bdz} := B'/2 \qquad \text{and} \qquad \tau_{\bdu} := B'/(2\sigma), \]
and let $\calD$ be the distribution over $\{0, \pm 1\}$ with $\Pr[0] = 1/2$
and $\Pr[1] = \Pr[-1] = 1/4$.
The gap requirement guarantees $\tau_{\bdu} \ge 13.78 \sqrt{m} B$, so an honest
prover passes the rejection check on $\bdu$ below except with negligible
probability: by a Hoeffding bound, each coordinate of $\bdR \bda$ exceeds
$\tau_{\bdu}$ with probability at most $2e^{-\tau_{\bdu}^2/(2 m B^2)} \le
2e^{-94.9}$, so all $\sigma$ coordinates pass except with probability
$2^{-128}$, which also bounds what the rejection on $\bdu$ reveals about $\bda$.

\begin{myenumerate2}  
    \item The prover first sends $\commit(\bda)$ to the verifier.
    \item The prover chooses a uniform length-$\sigma$ vector $\bdy \xleftarrow{\$} [ \pm \lceil B'/2 \cdot (1+1/\sigma)\rceil   ]^{\sigma} $, and sends $\commit(\bdy)$ to the verifier.
    \item The verifier chooses $\bdR \leftarrow \calD^{\sigma \times m}$ and sends it to the prover. 
    \item The prover computes $\bdu := \bdR \cdot \bda$ and $\bdz = \bdu + \bdy$. It restarts the protocol from Step 2 if either $\| \bdu \|_\infty > \tau_{\bdu}$ or $\| \bdz \|_\infty > \tau_{\bdz}$.
    \item The prover sends $\bdz$ to the verifier. The verifier rejects
    immediately if $\| \bdz \|_\infty > \tau_{\bdz}$; this norm check is what
    enforces smallness, and the remaining steps only establish that $\bdz$ is
    consistent with the committed $\bda$ and $\bdy$.
    \item The verifier chooses a random $r$ and sends $r$ to the prover.
    \item The prover and the verifier run an inner-product proof that 
        \[ \langle \bdR^\top \bdr , \bda \rangle + \langle  \bdr , \bdy \rangle = \langle  \bdR^\top \bdr | \bdr,  \bda |\bdy \rangle = \langle \bdz, \bdr \rangle,\]
        where $\bdr = (r^0, r^1, \ldots, r^{\sigma-1})$. The commitment for
        this proof is assembled homomorphically from $\commit(\bda)$ and
        $\commit(\bdy)$, which binds the relation to the vectors committed in
        Steps 1 and 2. (Non-interactively, $\bdR$ and $r$ are derived via
        Fiat-Shamir: $\bdR$ from the iteration's public parameters, $\commit(\bda)$ and $\commit(\bdy)$, and
        $r$ from the transcript including $\bdz$.)
\end{myenumerate2}
Note that $\langle \bdz, \bdr \rangle$ is a public value. The last step is essentially a length-$(m+\sigma)$ inner product proof. 

The thresholds above are the standard parameters for rejection sampling with a
uniform mask: the mask amplitude is $\tau_{\bdz}(1+1/\sigma) = \tau_{\bdz} +
\tau_{\bdu}$, so conditioned on acceptance, $\bdz$ is uniform on $[\pm
\tau_{\bdz}]^{\sigma}$ and independent of $\bda$ (zero knowledge), and each
attempt is accepted with probability $\left(\sigma/(\sigma+1)\right)^{\sigma}
\approx 1/e$, i.e., expected $e \approx 2.72$ attempts.

There are two properties of this approximate proof: correctness and soundness.
Correctness means that if a client has a vector of $L_\infty$ norm smaller than $B$ then it should prove it except with negligible failure probability.
Soundness means that if a client has a vector with $L_\infty$ norm larger than $B'$, then it fails the approximate proof with overwhelming probability. Concretely, for fixed $\bda$ with $\|\bda\|_\infty > B'$ and fixed $\bdy$, each row of $\bdR$ lets $\bdz$ pass the check $\|\bdz\|_\infty \le \tau_{\bdz}$ with probability at most $1/2$~\cite[Lemma 2.5]{lyubashevsky2021shorter}, so the proof passes with probability at most $2^{-\sigma}$; with Fiat-Shamir, a prover making $Q$ hash queries passes with probability at most about $Q \cdot 2^{-\sigma}$. As in~\cite{GHL22}, we take $\sigma = 128$. Note that this soundness does not mean that a client will fail the proof when the vector has $L_\infty$ norm between $B$ and $B'$; and this is exactly why we cannot merely use approximate proof for our proof of $L_\infty$ norm.

\subsection{Full protocol for proof of encryption}\label{app:full-enc-proof}
We now describe proof protocol $\Pi_{\text{enc}}$, which is proof of Regev's
encryption with bounded norms for error and input.

Recall the constraints that client $i$ wishes to prove are
$$
\begin{aligned}
\CSEnc:
\{
    & \zkio: (\bdy_i,\commit(\lwesecret_i), \commit(\bdx_i), \commit(\bde_i^{\mathrm{hi}})), \\
    & \zkstate: \bdy_i=\bdA \lwesecret_i+\bde_i + \Delta \bdx_i, \\
    & \qquad \| \bdx_i \|_2 \le B_{\bdx}(L_2),\\
    & \qquad \| \bdx_i \|_\infty \le B_{\bdx}(L_\infty), \| \bde_i \|_\infty \le B_{\bde}(L_\infty),\\
    & \witness: (\lwesecret_i, \bdx_i, \bde_i)\} \\
\end{aligned}
$$ 
This can be proven using the techniques we presented in Section~\ref{s:robust}
and~\ref{s:proof-norms}. Below we omit index $i$ for simplicity. Assuming the
prover already commits to $\bdx,\bde^{\mathrm{hi}},\lwesecret$, where $\bde =
2^{h}\bde^{\mathrm{hi}} + \bde^{\mathrm{lo}}$ (\S\ref{s:proof-norms}), and has
fixed the public ciphertext $\bdy$.
\begin{enumerate} 
    \item The prover invokes the approximate proof (\appref{app:approx}) on
    $\bde^{\mathrm{lo}} = \bdy - \bdA\lwesecret - 2^{h}\bde^{\mathrm{hi}} -
    \Delta\bdx$, which is not committed: with its projection matrix $\bdR$, mask
    $\bdy'$, masked projection $\bdz'$ and $\vec{r}:=(r^0, \ldots,
    r^{\sigma-1})$, its linear proof is on
    \[ \langle \bdz', \vec{r}\rangle - \langle \bdR^\top\vec{r}, \bdy\rangle =
    \langle -\bdA^\top \bdR^\top\vec{r} \,|\, -2^{h}\bdR^\top\vec{r} \,|\,
    -\Delta \bdR^\top\vec{r} \,|\, \vec{r},\ \lwesecret \,|\,
    \bde^{\mathrm{hi}} \,|\, \bdx \,|\, \bdy' \rangle, \]
    whose commitment is assembled homomorphically from $\commit(\lwesecret)$,
    $\commit(\bde^{\mathrm{hi}})$, $\commit(\bdx)$ and $\commit(\bdy')$ as in
    \S\ref{s:proof-norms}. This proves $\bdy=\bdA\lwesecret+\bde+\Delta\bdx$
    and $\|\bde^{\mathrm{lo}}\|_\infty \le \gamma (2^{h-1}+1)$.

    \item The prover invokes proof of $L_2$ norm (described in
    \S\ref{s:proof-norms}) on $\bdx$.
    \item The prover invokes proof of $L_\infty$ norm (described in
    \S\ref{s:proof-norms}) on $\bde^{\mathrm{hi}}$ and $\bdx$, with $\commit(\vec{z})$
    assembled from $\commit(\bde^{\mathrm{hi}})$ and $\commit(\bdx)$, respectively.
    
\end{enumerate}

\paragraph{Implementation of the proof of $L_\infty$ norm.}
Our implementation uses the decomposition into three numbers
(\S\ref{s:proof-norms}), so $\vec{z}$ has $4\ell$ entries. The approximate
proof on $\vec{z}$ (\appref{app:approx}) commits to its mask $\bdy$ on the
generators right after these $4\ell$ entries, before $r$ is sampled, and
$\vec{r}'$ is set to zero on the entries holding $\bdy$. The linear proof for
Equation~\ref{eq:extra-ip} and the linear proof in the approximate proof are
then combined into one linear proof, by taking a random linear combination of
the two relations with a challenge derived by Fiat-Shamir after both are
fixed; its commitment is assembled homomorphically as in
\S\ref{s:proof-norms}. The public vector in Equation~\ref{eq:extra-ip} is
non-zero on every entry of $\vec{t}$ and, since $\vec{r}'$ is zero there, zero
on the entries holding $\bdy$; hence the combined proof forces the
corresponding entries of $\vec{t}$ to be zero, and $\bdy$ does not enter the
relation in Equation~\ref{eq:Linf-ip}. This removes the separate
length-$(4\ell+\sigma)$ linear proof of the approximate proof: the combined
linear proof has length $2(4\ell+\sigma)$, instead of $8\ell$ and
$4\ell+\sigma$ for the two proofs. For the $L_2$ proof, the implementation pads
$\bdv$ with zeros to a power-of-two length $L \ge \ell+4$, so its consistency
proof and approximate proof have length $2L$ (the approximate proof also covers
the padding, which the wraparound bound $\sqrt{q/(2L)}$ accounts for).

\section{Deferred material for baseline}\label{app:eval}

\subsection{Cost overview of ACORN-detect}\label{app:baselines:acorn-detect}
To understand how ACORN-robust works we first present ACORN-detect. 
In ACORN-detect protocol, the server can detect if a client cheats but the protocol does not have guaranteed output delivery. We outline the protocol below and briefly analyze its cost.

We start with the protocol (without input validation) in Bell et al.~\cite{bell20secure} which is also a base protocol for ACORN. 
Initially, the server establishes a public graph on all $n$ clients where each client has $k=O(\log n)$ neighbors; let $N(i) \subset [n]$ denote the neighbors of $i$. 
Each pair of clients establish pairwise secrets $p_{ij}$.
Each client $i$ generates a random PRG seed $z_i$ and masks the input $\bdx_i$ as 
\[\bdy_i = \bdx_i + \bdr_i , \]
where the mask $\bdr_i$ is defined as
\[\bdr_i = \sum_{i<j, j\in N(i)} \PRG(p_{ij}) - \sum_{i>j,j\in N(i)} \PRG(p_{ij}) + \PRG(z_i). \]
The client sends $\bdy_i$ to the server.
Note that $z_i$ is for ensuring privacy when handling dropouts; see more details in Bell et al.~\cite{bell20secure}. 
We skip the rest of details of the protocol here, but their key feature is that for any online set $\mathcal{O}\subset [n]$, the server eventually gets $\bdr := \sum_{i\in \mathcal{O}} \bdr_i$ so that it can remove $\bdr$ from $\bdy := \sum_{i\in \mathcal{O}} \bdy_i$ and obtain the desired output $\sum_{i\in\mathcal{O}} \bdx_i$.  

To achieve input validity, they added the following steps to the above protocol. 
Each client $i$ computes the commitment to $\bdx_i$ and the commitment to the aggregated mask $\bdr_i$ and sends them together with the masked vector $\bdy_i = \bdx_i + \bdr_i$.  
Then the client proves that 
\begin{myitemize2}
    \item $\bdx_i$ has valid $L_2, L_\infty$ norm (same as Section~\ref{s:proof-norms});
    \item It added $\bdr_i$ to $\bdx_i$ correctly (which can be done using a linear proof). 
\end{myitemize2}
Recall that the server learns $\bdr := \sum_{i\in \mathcal{O}} \bdr_{i}$.  %
Next, the clients and the server run a distributed key correctness (DKC) protocol to check if the server obtains $\bdr := \sum_{i\in \mathcal{O}} \bdr_{i}$ where the $\bdr_i$'s are indeed consistent with the commitments that the clients sent in the first place. 

\begin{remark}
    When the PRG is instantiated with homomorphic PRG (e.g., RLWE-based PRG), the client can optimize its computation by first computing the sum of the seeds and then expanding the aggregated seeds with PRG. A trade-off is that the masking here is not simply $\bdx_i+\bdr_i$: since the PRG output is defined over polynomial rings, the input $\bdx_i$ should be interpreted as polynomials when added to $\bdr_i$ and this requires non-trivial encoding of $\bdx_i$ (see Equation 5 in ACORN~\cite{bell22acorn}). As a result, the client also needs to prove it performs the encoding correctly.
\end{remark}

\paragraph{Cost.}
Each client computes two vector commitments to length $\ell$ vectors $\bdx_i, \bdr_i$. 
For the DKC protocol, the client performs a constant number of elliptic curve scalar multiplications, and the server performs $3n$ of them.

\subsection{Cost overview of ACORN-robust}\label{app:baselines:acorn-robust}
ACORN-robust is similar to ACORN-detect but with the following differences:
\begin{myitemize2}
    \item The pairwise secrets are established differently (see details below);
    \item When the server fails verification in the DKC protocol, it invokes an $O(\log n)$-round bad message resolution protocol with all the clients to remove the malicious clients' contribution from the sum. 
\end{myitemize2}

Suppose the server establishes a public graph on all $n$ clients where each client has $k=O(\log n)$ neighbors.
First, each client $i$ generates $k$ seeds $s_{i, j}$ for neighbor $j$, and sends them to the neighbors; client $i$ additionally generates (deterministic) commitments to the seeds, namely $s_{i,j} \cdot G$, which are sent to the server. 
Next, clients exchange the seeds with their neighbors: a client $i$ neighboring with client $j$ will send $s_{i,j}$ and receive $s_{j,i}$, and vice versa. Client $i$ and $j$ then establish pairwise secret $p_{i j} = s_{i,j} + s_{j,i}$; this $p_{ij}$ will be used for pairwise-masking the input vector. 

Each client $i$ then Shamir-shares $s_{i,j}$ and sends the Feldman commitments to the sharing of $s_{i,j}$ (commitments to the coefficients of the sharing polynomial) to the server. The server checks if the Feldman commitments match the commitment $s_{i,j} \cdot G$. If not, the server disqualifies client $i$; if it matches, the server computes $s_{i,j}^{(k)}\cdot G$ from Feldman commitments, where $s_{i,j}^{(k)}$ is the share meant for the $k$-th neighbor of client $i$. Then the server sends $s_{i,j}^{(k)}\cdot G$ to the corresponding client. The recipient client checks if the decrypted share $s_{i,j}^{(k)}$ matches the commitment $s_{i,j}^{(k)}\cdot G$. Then the server and clients invoke an $O(\log n)$-round bad message resolution protocol to form a set of clients whose pairwise masks can be canceled out.

There are two costly parts of ACORN-robust: 1) the obvious complexity of the logarithmic number of rounds between the server and all the clients; 2) the server needs to verify $O(n \log n)$ Feldman commitments of sharing of degree $\log n$. 
Concretely, with $n=1000$ clients and $40$ neighbors per client, as in \S\ref{s:eval}, the server needs to perform $40^2 \cdot 1000=1,600,000$ elliptic curve scalar multiplications, which takes about 40 seconds (\S\ref{s:eval-bench}); this cannot be trivially optimized with multi-exponentiation because the server needs to identify the malicious clients.

\section{Security proof of \sys{}}\label{app:security}

We give our full protocol description in Figures~\ref{fig:full} and \ref{fig:full2}.

\newcommand{\cA}{\mathcal{A}}
\newcommand{\cT}{\mathcal{T}}
\newcommand{\cF}{\mathcal{F}}
\newcommand{\cX}{\mathcal{X}}
\newcommand{\term}[1]{\mathsf{#1}}
\newcommand{\Ideal}{\term{Ideal}}
\newcommand{\Real}{\term{Real}}

\subsection{Proof of Theorem~\ref{thm:main}}\label{app:proof:main}
We follow the proof of security similar to that of ACORN-robust~\cite{bell22acorn}. Their protocol guarantees the privacy of honest clients only with a semi-honest server. This is an artifact of their protocol where the server is empowered to recover the masks---both the self-masks and pairwise masks---for misbehaving clients to then remove their inputs. In other words, the server is capable of recovering the actual inputs of malicious clients. Consequently, a malicious server could claim honest clients to be malicious and thereby recover the inputs of these clients. Our protocol uses a single mask per client, which the server does not recover, but its complaints rely on a semi-honest server in a similar way (Remark~\ref{sub:mal}); like theirs, our privacy proof assumes a semi-honest server.

\newcommand{\tSim}{\mathsf{Sim}}
\newcommand{\bdX}{\mathbf{X}}
\newcommand{\bdK}{\mathbf{K}}
\newcommand{\bfx}{\bdx}
\newcommand{\bfA}{\bdA}
\newcommand{\bfy}{\bdy}
\newcommand{\bfe}{\bde}
\newcommand{\bdf}{\mathbf{f}}
\newcommand{\bbZ}{\mathbb{Z}}
\newcommand{\cD}{\mathcal{D}}
\newcommand{\getsr}{\xleftarrow{\$}}
Our proof methodology relies on the standard simulation-based proof, %
where we show that every adversary attacking our protocol can be simulated by an adversary $\tSim$ in an ideal world where the functionality $\mathcal{F}$ (Fig.\ref{fig:ideal-sum}). In the following, we first prove privacy against an adversary that corrupts the server semi-honestly and $\eta n$ clients and $\eta_\comset C$ \coms maliciously; then we prove robustness assuming the adversary corrupting $\eta n$ clients and $\eta_\comset C$ \coms but not the server (recall our threat model in \S\ref{s:pb}). 

The challenge in the simulation is the ability of $\tSim$ to generate a valid distribution for the honest clients' inputs, even without knowing their keys. To this end, we will show that $\tSim$, when only given the sum of the user inputs $\bdX=\sum_{i=1}^n \bdx_i$ (below, $[n]$ stands for the honest clients whose ciphertexts are summed), can simulate the expected leakage for the server which includes $n$ ciphertexts, the sum of the $n$ keys $\bdK=\sum_{i=1}^{n}\bdk_i$, and such that the sum of the $n$ ciphertexts, when decrypted with $\bdK$, correctly decrypts to $\bdX$.

Before we detail the definition of $\tSim$ and prove its security, we present an assumption that we will use later.

\begin{definition}[A variant of Hint-LWE~\cite{lee18instant, cheon21lattice}]
\label{def:hint-lwe}
Consider integers $\lambda, m, q$
and a probability distribution $\chi'$ on $\bbZ_q$, typically taken to be a normal
distribution that has been discretized. Then, the Hint-LWE assumption states
that for all PPT adversaries $\mathcal{A}$, there exists a negligible function $\mathsf{negl}$ such that: 
\[
\Pr\left[b=b'~~
\begin{array}{|c}
      \bfA\gets \mathcal{D}_{\bfA},\bdk\getsr \bbZ_q^{\lambda},\bfe\getsr \chi'^m\\
      \bdr\getsr\bbZ_q^{\lambda},\bdf\getsr\chi'^m\\
      \bfy_0:=\bfA \bdk+\bfe,
      \bfy_1\getsr\bbZ_q^m,
      b\getsr\{0,1\}\\
      b'\getsr\cA(\bfA,(\bfy_b,\bdk+\bdr,\bde+\bdf))
\end{array}
\right]=\frac{1}{2} + \mathsf{negl}(\kappa)
\]
where $\kappa$ is the security parameter and $\mathcal{D}_{\bfA}$ is the
distribution of the public matrix in $\bbZ_q^{m\times\lambda}$ (uniform, or
sparse as in \S\ref{s:impl}; the matrix of an iteration is expanded from the
setup string and the iteration number, and is modeled as a sample from
$\mathcal{D}_{\bfA}$). The simulation below does not use the distribution of
$\bfA$.

\end{definition}
Intuitively, Hint-LWE assumption says that $\bdy_0$ looks pseudorandom to an adversary, even when given some randomized leakage on the secret and the error vectors. Lee~\etal~\cite[Theorem 3]{lee18instant} show that solving Hint-LWE is no easier than solving LWE problem, for any secret distribution (for discrete Gaussians, by the proof of~\cite[Theorem 1]{C:KLSS23b}, whose factor 2 is not needed when $\sigma'$ is far above the smoothing parameter). For a secure LWE instance $(\lambda, m, q)$ with discrete Gaussian error of standard deviation $\sigma$, the corresponding Hint-LWE instance $(\lambda, m, q, \chi')$, where $\chi'$ is a discrete Gaussian distribution with standard deviation $\sigma'$, is secure when $\sigma'=\sqrt{2}\sigma$. The Regev error is $\bde=\bde_1+\bde_2$ with $\bde_1,\bde_2\getsr\chi'^m$ (i.e., $\chi=\chi'+\chi'$), of standard deviation $\sqrt{2}\sigma'=2\sigma$. This gives us the real distribution $\cD_R$, with the error term re-written and the last ciphertext modified. 
{\setlength{\arraycolsep}{2pt}\begin{gather*}
    \left\{
\begin{array}{c|c}
  \bdK=\sum_{i=1}^{n} \bdk_i \bmod q   & \forall i\in[n], \bdk_i\getsr\bbZ_q^\lambda,\bde_i,\bdf_i\getsr\chi'^m \\
 \bdy_1,\ldots,\bdy_n  &  \forall i\in[n-1], \bdy_i=\bdA\cdot \bdk_i +\bde_i+\bdf_i+\Delta\bdx_i\\
 & \bdy_n=\bdA \bdK-\sum_{i=1}^{n-1} \bdy_i +\sum_{i=1}^n (\bde_i+\bdf_i)+\Delta \bdX
\end{array}
\right\}
\end{gather*}}

We now define $\tSim(\bdA,\bdX)$:
\begin{algorithmic}
\State \underline{$\tSim(\bdA,\bdX)$}
\State Sample $\bdu_1,\ldots,\bdu_{n-1}\getsr\bbZ_q^m$
\State Sample $\bdk_1,\ldots,\bdk_n\getsr\bbZ_q^\lambda$
\State Sample $\bde_1,\ldots,\bde_n \getsr\chi'^m$
\State Sample $\bdf_1,\ldots,\bdf_n \getsr\chi'^m$
\State Set $\bdK:=\sum_{i=1}^n \bdk_i \bmod q$
    \State Set $\bdu_n=\bdA\cdot \bdK-\sum_{i=1}^{n-1} \bdu_i +\sum_{i=1}^n (\bde_i+\bdf_i)+\Delta\cdot \bdX$
    \State Return $\bdK,\bdu_1,\ldots,\bdu_n$
\end{algorithmic}

In other words, the simulated distribution, $\mathcal{D}_{\tSim}$, is:
{\setlength{\arraycolsep}{2pt}\begin{gather*}
\left\{
\begin{array}{c|c}
  \bdK=\sum_{i=1}^{n} \bdk_i \bmod q   & \forall~i\in[n]~\bdk_i\getsr\bbZ_q^\lambda,\bde_i,\bdf_i\getsr\chi'^m \\
 \bdu_1,\ldots,\bdu_n  &  \forall~i\in[n-1]~ \bdu_i\getsr\bbZ_q^m\\
 & \bdu_n=\bdA \bdK-\sum_{i=1}^{n-1} \bdu_i +\sum_{i=1}^n (\bde_i+\bdf_i)+\Delta \bdX
\end{array}
\right\}
\end{gather*}}

We will now prove that $\cD_{R}$ is indistinguishable from $\cD_{\tSim}$ through a sequence of hybrids. 

\begin{itemize}
    \item Hybrid 0: This is $\cD_R$. 
    \item Hybrid 1: In this hybrid, we will replace the real ciphertext $\bdy_1$ with a modified one. In other words, we set: 
{\setlength{\arraycolsep}{1pt}\begin{gather*}
\left\{
\begin{array}{c|c}
  \bdK & \forall i\in[n], \bdk_i\getsr\bbZ_q^\lambda,\bde_i,\bdf_i\getsr\chi'^m, {\color{blue}\bdu_1'\getsr\bbZ_q^m} \\
 {\color{blue}\bdy_1=\bdu_1'+\bdf_n+\Delta\bdx_1} &  \forall i\in[2,n-1], \bdy_i=\bdA\cdot \bdk_i +(\bde_i+\bdf_i)+\Delta\bdx_i\\
\{\bdy_i\}_{i=2}^{n}  & \bdy_n=\bdA \bdK-\sum_{i=1}^{n-1} \bdy_i +\sum_{i=1}^n (\bde_i+\bdf_i)+\Delta \bdX
\end{array}
\right\}
\end{gather*}}
Now, we will show that if there exists an adversary $\mathcal{B}$ that can distinguish between Hybrid 0 and 1, then
we can define an adversary $\mathcal{A}$ who can distinguish the two ensembles in the Hint-LWE Assumption. Let us define
$\mathcal{A}$ now. 

\begin{algorithmic}
    \State \underline{$\mathcal{A}(\bdA,\bdy^\ast,\bdk^\ast=\bdk+\bdr\bmod q,\bde^\ast=\bde+\bdf$)}
    \State Sample $\bdk_2,\ldots,\bdk_{n-1}\getsr\bbZ_q^\lambda$
    \State Sample $\bde_2,\ldots,\bde_{n}\getsr\chi'^m$
    \State Sample $\bdf_2,\ldots,\bdf_{n}\getsr\chi'^m$
    \State Set $\bdK=\sum_{i=2}^{n-1} \bdk_i + \bdk^\ast\bmod q$ \Comment{implicitly, $\bdk_n:=\bdr$}
    \State $\forall~i\in \{2,\ldots,n-1\},~\bdy_i=\bdA\bdk_i+\bde_i+\bdf_i+\Delta\bdx_i$
    \State Set $\bdy_1=\bdy^\ast+\bdf_n+\Delta\bdx_1$
    \State Set $\bdy_n:=\bdA\bdK-\sum_{i=1}^{n-1} \bdy_i+\bde^\ast+\sum_{i=2}^{n} (\bde_i+\bdf_i)+\Delta\cdot \bdX$
    \State Run $b'\getsr\mathcal{B}(\bdK,\bdy_1,\ldots,\bdy_n)$
    \State \textbf{return} $b'$
\end{algorithmic}
We need to argue that the reduction correctly simulates the two hybrids, based on the choice of $y^\ast$. 

\begin{itemize}
    \item If $\bdy^\ast=\bdA\bdk+\bde$, then $\bdy_1$ is a valid encryption of $\bdx_1$ with key $\bdk$ and error $(\bde+\bdf_n)$. Further, it is easy to verify that $\bdy_n$ satisfies the definition present in Hybrid 0. 
    \item If $\bdy^\ast=\bdu$ for some random $\bdu$. Then, we get that $\bdy_n$ is of the prescribed format, while also guaranteeing that $\bdy_1$ is generated as expected. 
\end{itemize}
\item Hybrid 2: In this hybrid, we will replace $\bdy_1$ with $\bdy_1$ that is sampled uniformly at random. 

{\begin{gather*}
\left\{
\begin{array}{c|c}
  \bdK & \forall~i\in[n]~\bdk_i\getsr\bbZ_q^\lambda,\bde_i,\bdf_i\getsr\chi'^m, {\color{blue}\bdu_1\getsr\bbZ_q^m} \\
 {\color{blue}\bdu_1} &  \forall~i\in[2,n-1]~ \bdy_i=\bdA\cdot \bdk_i +(\bde_i+\bdf_i)+\Delta\bdx_i\\
\{\bdy_i\}_{i=2}^{n}  & \bdy_n=\bdA \bdK-\bdu_1-\sum_{i=2}^{n-1} \bdy_i +\sum_{i=1}^n (\bde_i+\bdf_i)+\Delta \bdX
\end{array}
\right\}
\end{gather*}}
Hybrid 1 and Hybrid 2 are identically distributed because $\bdu_1'$ is uniformly sampled and masks the added error share in $\bdy_1$ of Hybrid 1.
\end{itemize}
In Hybrids 3 and 4, we replace $\bdy_2$ with a random element $\bdu_2$, by using a similar logic. Therefore, in Hybrid $2n-2$, the distribution will resemble $\cD_{\tSim}$. This concludes the proof of simulatability. 

\paragraph{Privacy against a semi-honest server colluding with malicious clients.}
Here the adversary $\Adv$ controls the server, which follows the protocol but shares its view with $\Adv$, and maliciously controls a set $\mathsf{Cor}\subseteq[n]$ of $\eta n$ clients and a set $\mathsf{Cor}_\comset\subseteq\comset$ of $\eta_\comset C$ \coms, which may deviate arbitrarily from the protocol; corrupted clients may send their Round 1 messages after seeing the honest ones through the server. In $\mathcal{F}$, honest parties have no output and the \coms have no input; the server's output is part of $\Adv$'s view, and its correctness is the subject of Definition~\ref{def:robustness}. $\tSim$ receives $\mathcal{O}$ with these times (Figure~\ref{fig:ideal-sum}) before the aggregation starts. Let $\mathcal{H} := [n]\setminus(\mathsf{Cor}\cup\mathcal{O})$ be the honest clients whose Round 1 messages reach the server in time, and $\mathcal{L}\subseteq\mathcal{O}$ those whose Round 1 messages reach it only after its waiting time. Of the semi-honest server, the proof uses that it forwards to \com $j$ exactly client $i$'s $\mathsf{ct}_{i,j}$ and the commitment $\commit_{K_j}(\share_i^{(j)})$ bound by its $\pi_{\text{shares}}$, under $i$'s index; that it sends the same set $\mathcal{B}$ to all \coms, ignores messages that arrive after its waiting time, and accepts only complaints that verify; and that it uses an ID $t$ that is unique among all aggregations under the \coms' keys and the clients' IDs, fixed across aggregations, as indices (Figure~\ref{fig:full}), so that each label $(t,i,j)$ and each proof context belongs to one client. (If an aggregation were repeated under the same $t$, e.g., by a server that restarts or another training process that uses the same \coms, with another client at index $i$, that client could submit $i$'s earlier ciphertexts, valid under $(t,i,j)$, and obtain their keys from the complaints.) Channels are authenticated, so corrupted parties cannot send messages in the name of honest ones, and \coms without a valid registration are removed from $\comset$ (Figure~\ref{fig:full}).

The simulator $\tSim$ runs $\Adv$, plays the honest clients and \coms, and executes the server's steps as the protocol prescribes. Everything derived from the corrupted parties' messages (whether their proofs verify, the complaints about them, their shares in the \coms' sums, the decoding and the decrypted output) is thus computed from $\Adv$'s actual messages in every hybrid below, and the hybrids only change the honest parties' messages. In particular, this proof never extracts the corrupted clients' inputs; knowledge soundness is used only for robustness.

We use two facts. (a) $\mathcal{S}_3\cap([n]\setminus\mathsf{Cor}) = \mathcal{H}$, since the honest clients' inputs lie in $\mathcal{Z}$ (Theorem~\ref{thm:main}~(iv)): the message of a client in $\mathcal{H}$ is complete and its proofs verify, so it is in $\calX_2$, and a complaint about it is valid only if $\mathsf{ct}_{i,j}$ is invalid under the label $(t,i,j)$ or does not decrypt to an opening of $\commit_{K_j}(\share_i^{(j)})$. Neither holds: the key $u^{x_j}$ of a valid ciphertext is determined by $PK_j$ and $u$, and by the soundness of the Chaum-Pedersen proof no \com can prove another key, even for a key it registered itself. We show (a) in Hybrid 0; as the event is efficiently checkable from $\Adv$'s view, it carries over to the next hybrids by their indistinguishability. So $\tSim$ sends $\mathcal{E}=\emptyset$ to $\mathcal{F}$, which returns $\bdx_{\mathcal{H}} := \sum_{i\in\mathcal{H}}\bdx_i$. (b) What the honest \coms reveal about the honest clients' ciphertexts to them is a fixed function of their plaintexts. An honest \com decrypts the shares of the clients in $\mathcal{H}$ correctly and they open the commitments the server forwards, so it never complains about them; these shares enter only its Round 3 sum over $\mathcal{S}_3$, whose honest part is the sum over $\mathcal{H}$ by (a), whatever $\Adv$ does. The keys revealed in complaints are those of ciphertexts forwarded for corrupted clients, under labels $(t,i,j)$ with $i\in\mathsf{Cor}$. So the ciphertexts from honest clients to honest \coms are never decrypted for $\Adv$, and what is revealed about them is determined by $\mathcal{H}$ alone. Chosen-ciphertext security therefore suffices: unlike protocols in which a malicious server chooses which ciphertexts to open after seeing them~\cite[App.~A]{GKKTW26}, selective-opening security is not needed. This relies on the server forwarding each client's own commitments (Remark~\ref{sub:mal}).

\begin{itemize}
    \item Hybrid 0: This is the real execution of the protocol.
    \item Hybrid 1: $\tSim$ knows all the honest parties' inputs and keys, runs them as in the protocol, and programs the random oracle as needed. This is identical to Hybrid 0.
    \item Hybrid 2: $\tSim$ produces the proofs $\pi_{\text{enc}}$ and $\pi_{\text{shares}}$ of the clients in $\mathcal{H}\cup\mathcal{L}$ with the zero-knowledge simulators, programming the random oracle at points that $\Adv$ has queried only with negligible probability; the random oracles of the proofs and of $\AsymEnc$ are domain-separated. The corrupted clients' proofs are still verified as in the protocol. Indistinguishable by zero knowledge. The simulators do not use the witnesses, so they also run on the false statements of Hybrid 7.
    \item Hybrid 3: The honest \coms no longer decrypt the ciphertexts of the clients in $\mathcal{H}$: they take the shares that $\tSim$ generated and include those clients in $\mathcal{V}_j$ without complaint (they never receive the late ciphertexts of the clients in $\mathcal{L}$). They still check and decrypt every ciphertext forwarded for a corrupted client. This is identical to Hybrid 2, where these decryptions give exactly those shares, which open their commitments.
    \item Hybrid 4: $\tSim$ replaces every $\mathsf{ct}_{i,j}$ with $i\in\mathcal{H}\cup\mathcal{L}$ and honest $j$ by an encryption of $0$ under the same label. Indistinguishable by Theorem~\ref{thm:main}~(iii), with a hybrid over these ciphertexts. By (b), the reduction obtains everything an honest \com reveals or uses about the other ciphertexts from its oracles: their validity (a public check), the corrupted clients' shares in its Round 3 sum, and, in a complaint, the key $u^{x_j}$ with its Chaum-Pedersen proof, which is a decryption share in the model of~\cite{EC:ShoGen98}; complaints in other iterations are oracle queries too, as labels contain $t$. For several aggregations under the same keys (Theorem~\ref{thm:main}), the challenge ciphertexts are those from honest clients to honest \coms in every aggregation, and in the other aggregations the reduction runs the honest clients itself. It computes the honest part of the Round 3 sums from the shares it chose, and simulates the honest \coms' registration proofs.
    \item Hybrid 5: $\tSim$ replaces the commitments $\commit_{\bold{F}}(\lwesecret_i)$, $\commit_{\bold{G}}(\bde_i^{\mathrm{hi}})$, $\commit_{\bold{H}}(\bdx_i)$ of the clients in $\mathcal{H}\cup\mathcal{L}$, and their share commitments for honest \coms, by commitments to $0$, keeping the share commitments for corrupted \coms, which can decrypt their openings. The randomness of these commitments appears nowhere else since Hybrid 4 (their shares still enter the Round 3 sums until Hybrid 6), and the proofs are simulated, so this is identical by the perfect hiding of Pedersen commitments.
    \item Hybrid 6: The sharings of the clients in $\mathcal{H}\cup\mathcal{L}$ now reach $\Adv$ only through the shares for the corrupted \coms (with their commitments and ciphertexts) and, for $\mathcal{H}$, the honest \coms' Round 3 sums. $\tSim$ samples the shares of each such $\lwesecret_i$ for the corrupted \coms uniformly and recomputes their commitments and ciphertexts from them with fresh randomness. It computes the honest \coms' sums from a uniformly random polynomial of degree $d$ that takes the values $\bdK_{\mathcal{H}} := \sum_{i\in\mathcal{H}}\lwesecret_i$ at $0,\ldots,-\lambda+1$ and, at each corrupted \com, the sum over $\mathcal{H}$ of the shares sampled for it, adding as in the protocol the corrupted clients' shares that the honest \coms decrypt. This is identical given (a): since $|\mathsf{Cor}_\comset| \le \eta_\comset C < d-\lambda+1$, the shares of a degree-$d$ packed sharing for the corrupted \coms are uniform and independent of the secrets, and the sum of independent sharings conditioned on these values is uniform on the remaining affine space. The LWE secrets of the clients in $\mathcal{H}\cup\mathcal{L}$ now enter $\Adv$'s view only through $\bdK_{\mathcal{H}}$ and the ciphertexts $\bdy_i$.
    \item Hybrid 7: $\tSim$ computes $\bdK_{\mathcal{H}}$ and $\{\bdy_i\}_{i\in\mathcal{H}}$ with $\tSim(\bdA, \bdx_{\mathcal{H}})$ above, and samples $\bdy_i$ for $i\in\mathcal{L}$ uniformly. Indistinguishable by the Hint-LWE assumption for $\mathcal{H}$, as shown above, and by LWE for $\mathcal{L}$: the reduction computes the rest of $\Adv$'s view, including the output, from these values and $\Adv$'s messages. (If $|\mathcal{H}|=1$, $\bdx_{\mathcal{H}}$ is that client's input and the hybrid is trivial.)
\end{itemize}
In Hybrid 7, $\tSim$ uses only $\mathcal{O}$ with these times and $\bdx_{\mathcal{H}}$, which it obtains from $\mathcal{F}$ with $\mathcal{E}=\emptyset$. This concludes the proof. 

\begin{remark}[On privacy of ACORN-robust]\label{rmk:acorn-robust-semi-honest}
    A critical artifact of ACORN-robust in~\cite{bell22acorn} is the loop-based resolution of malicious behavior. Specifically, the protocol relies on
    a looping process by which
    the server identifies some malicious clients
    in every round of communication. This is done
    by finding inconsistencies in the clients' communication. Unfortunately, once a misbehaving client is detected, the protocol must communicate with the parties to retrieve the self-mask \emph{and} the pairwise masks along each edge of the neighborhood graph. Consequently, the server receives all the information necessary to unmask the inputs. Therefore, a malicious server could conceivably claim an honest client to be a misbehaving client, thereby compromising the privacy of the inputs. This is acknowledged by the authors of \cite{bell22acorn}. However, a simple fix would be for the server to attach necessary proofs of malicious behavior but the communication involved in this process is higher. 
\end{remark}

\paragraph{Robustness.}
Now the server follows the protocol and does not collude with the adversary, which maliciously controls a set $\mathsf{Cor}$ of at most $\eta n$ clients and $\eta_\comset C$ \coms (privacy in this setting is a special case of the previous paragraph). We show that the server outputs $\sum_{i\in\mathcal{S}_3}\bdx_i$, where $\bdx_i$ is client $i$'s input for $i\notin\mathsf{Cor}$ and, for $i\in\mathsf{Cor}$, a vector in $\mathcal{Z}$ fixed by the adversary's Round 1 message, and that $\mathcal{S}_3$ contains every honest client whose Round 1 message reaches the server in time. Counting the corrupted clients outside $\mathcal{S}_3$ as inputs $\bm{0}\in\mathcal{Z}$, this is disruption resistance (Definition~\ref{def:disruption-resistance}), and robustness (Definition~\ref{def:robustness}) when $\mathcal{Z}$ is the set of all inputs; with no corrupted clients, it is dropout resilience (Definition~\ref{def:dropout-resilience}).
\begin{itemize}
    \item Honest clients are in $\mathcal{S}_3$: as in fact (a) above, the message of an honest client that reaches the server in time is in $\calX_2$, and no complaint about it verifies.
    \item Corrupted clients in $\mathcal{S}_3$ are well formed. Let $i\in\mathcal{S}_3\cap\mathsf{Cor}$. Its proofs verify, so by the knowledge soundness of $\pi_{\text{enc}}$ and the norm proofs there are $\lwesecret_i$, $\bdx_i\in\mathcal{Z}$ and $\bde_i$ with $\|\bde_i\|_\infty\le B_\bde$ and $\bdy_i = \bdA\lwesecret_i + \bde_i + \Delta\bdx_i$, where $\lwesecret_i$ is the vector committed in $\commit_{\bold{F}}(\lwesecret_i)$. By the knowledge soundness of $\pi_{\text{shares}}$ and Schwartz-Zippel over its weights $\bm{\gamma}$ (extracting from transcripts with independent weights, or in the algebraic group model), each $\commit_{K_j}(\share_i^{(j)})$ has no component on the generators $K_h$, $h\neq j$, or on $\bold{F}$, and its component on $K_j$ is the $j$-th share of a degree-$d$ packed sharing $f_i$ of $\lwesecret_i$. An honest \com $j$ that answers in Round 3 has $i\in\mathcal{V}_j$, i.e., it holds an opening of $\commit_{K_j}(\share_i^{(j)})$ on $(K_j, H)$ (otherwise its complaint would have removed $i$, or it would abstain), so by binding its share is $f_i(j)$. This does not rely on corrupt \coms being few: without the weights $\bm{\gamma}$, a single \com that drops out before checking its commitment would let a client shift the shares of all the others.
    \item The \coms' sums decode to $\sum_{i\in\mathcal{S}_3}\lwesecret_i$. The server sends the same $\mathcal{B}$ to all \coms, so the honest \coms that answer in Round 3 sum over the same $\mathcal{S}_3$, and their answers are evaluations of $F := \sum_{i\in\mathcal{S}_3} f_i$, a polynomial of degree $d$ that takes the values $\sum_{i\in\mathcal{S}_3}\lwesecret_i$ at $0,\ldots,-\lambda+1$. The honest \coms that drop out or abstain are erasures and the at most $\eta_\comset C$ corrupt \coms are errors, which is within the error-correction condition of \S\ref{s:security} (Theorem~\ref{thm:main}~(v)), so the server decodes $F$.
    \item Decryption is correct: $\sum_{i\in\mathcal{S}_3}\bdy_i - \bdA\sum_{i\in\mathcal{S}_3}\lwesecret_i = \sum_{i\in\mathcal{S}_3}\bde_i + \Delta\sum_{i\in\mathcal{S}_3}\bdx_i$, the error is smaller than $\Delta/2$ in every coordinate, and the sum of the inputs does not wrap modulo $p$ (Theorem~\ref{thm:main}~(iv)), so the server outputs $\sum_{i\in\mathcal{S}_3}\bdx_i$.
\end{itemize} 

\newcommand{\ignore}[1]{}

\subsection{A fix to ACORN-detect}\label{s:fix-acorn}
We clarify details related to counting rounds in experiments, point out an overlooked issue in ACORN-detect, and propose a patch. 

    ACORN-detect, as described in Figure 6
    of~\cite{bell22acorn},
    achieves input validation
    by integrating the distributed key correctness (DKC) protocol (as described in Figure 2 of~\cite{bell22acorn}) 
    and zero-knowledge proof  
    into the main secure aggregation protocol.
    The DKC protocol 
    is an interactive protocol
    which helps the server
    verify that the masks the server reconstructs
    is what the clients committed to 
    when sending the masked inputs to the server.
    In the protocol description of ACORN-detect in Figure 6, communication of the distributed key correctness protocol
    is embedded into the main protocol,
    thus there is no additional communication round incurred.
    However,
    it seems that the authors overlooked the assumption that
    the clients can drop offline in any round
    in the protocol execution
    when plugging the DKC protocol into ACORN-detect. 
    More specifically,
    the set of clients who participate in step 8
    in ACORN-detect
    (which contains step 3 of DKC)
    in which each client sends both the masked input and the commitment to the mask the user
    might be a superset of 
    the set of clients participating in Step 10
    (which contains step 5 of DKC)
    in which each client sends the server the information needed to verify the commitment of the mask
    if some clients drop offline between these two rounds.
    Note that the set $\mathcal{O}$
    of clients whose inputs are chosen to be included in the final result
    is determined when the server receives the masked input in step 9 of ACORN-detect
    and is not changed later.
    As a consequence, 
    in the last step of ACORN-detect (which contains step 6 of DKC),
    the server is not able to collect all the information needed for the key verification
    for the online set
    and the server will abort due to the verification failure even when all participants are honest,
    which breaks dropout resilience.
    This problem can be fixed by extracting
    step 4 and 5 of the DKC protocol from ACORN-detect as a separate round 
    between steps 8 and 9 of ACORN-detect
    rather than embedded in step 9 and 10 of ACORN-detect
    and determining the online set $\mathcal{O}$
    by who sends both the commitment of the masks and the information needed for the verification of the commitment.
    This fix introduces one extra round to ACORN-detect.

\fi

\end{document}